\pdfoutput=1
\documentclass[pdflatex,sn-nature]{sn-jnl}

\usepackage{graphicx}
\usepackage{multirow}
\usepackage{amsmath,amssymb,amsfonts,amsthm}
\usepackage{booktabs}
\usepackage[table,xcdraw]{xcolor}
\usepackage{colortbl}
\usepackage{tabularx}
\usepackage{bm}
\usepackage{algorithm}
\usepackage{algpseudocode}
\usepackage{tikz}
\usetikzlibrary{arrows.meta,positioning,shapes.geometric}
\usepackage{enumitem}
\usepackage{mathtools}
\usepackage{siunitx}
\usepackage{hyperref}

\newtheorem{remark}{Remark}

\begin{document}

\title[Cross-Temperature Defect Identification via Multi-Level Domain Alignment]{Cross-Temperature Defect Identification in Atomistic Simulations via Multi-Level Domain Alignment}

\author*[1]{\fnm{Yating} \sur{Fang}}\email{yating.fang@rutgers.edu}
\author[2]{\fnm{Jungmin} \sur{Kim}}
\author[3]{\fnm{Qian Qian} \sur{Zhao}}
\author[3]{\fnm{Pallavi} \sur{Biswas}}
\author[3]{\fnm{Joshua M.} \sur{Gonjon}}
\author[3]{\fnm{Ryan B.} \sur{Sills}}\email{ryan.sills@rutgers.edu}
\author[2]{\fnm{Ahmed} \sur{Aziz Ezzat}}\email{aa2085@rutgers.edu}

\affil*[1]{\orgdiv{Rutgers Artificial Intelligence and Data Science (RAD) Collaboratory}, \orgname{Rutgers University}, \orgaddress{\city{Piscataway}, \postcode{08854}, \state{NJ}, \country{USA}}}
\affil[2]{\orgdiv{Department of Industrial and Systems Engineering}, \orgname{Rutgers University}, \orgaddress{\city{Piscataway}, \postcode{08854}, \state{NJ}, \country{USA}}}
\affil[3]{\orgdiv{Department of Materials Science and Engineering}, \orgname{Rutgers University}, \orgaddress{\city{Piscataway}, \postcode{08854}, \state{NJ}, \country{USA}}}

\abstract{
Crystalline defects control how materials deform, transport mass, and accumulate radiation damage, yet near the melting point, where these processes are most active, thermal fluctuations distort local atomic environments until established structure-identification tools degrade sharply. We present a machine-learning framework that identifies point defects reliably up to the melting temperature. Scored against exhaustive Wigner--Seitz ground truth on vacancy- and self-interstitial-containing face-centered-cubic, body-centered-cubic, and hexagonal-close-packed iron systems, it localizes every interstitial and makes zero false detections in every vacancy system, with no high-temperature labels used in training. Deployed on a million-atom, 2.5\,ns trajectory, it sustains that fidelity at scale, resolving single vacancy hops and complete Frenkel-pair recombination, and applied to aluminum bicrystals it captures grain-boundary phase transformations in progress, distinguishing two nucleation modes. The framework works by treating the temperature gap as a label-transfer problem: trustworthy labels exist only in low-temperature reference structures, so it aligns the two regimes at three levels, denoising input coordinates, matching learned representations across temperature, and steering predictions toward the compact morphology of physical defect structures. Because no atom-wise truth exists at the target temperature, a label-free evaluation suite enables model assessment where ground truth cannot be constructed, completing a practical, label-efficient route to temperature-robust analysis of large-scale molecular dynamics.
}


\maketitle

\section{Introduction}
\label{sec:introduction}

Atomistic simulations are increasingly used to study a wide range of phenomena, including biological systems~\cite{gelpiMolecularDynamicsSimulations2015,venableMolecularDynamicsSimulations2019}, crystal nucleation~\cite{sossoCrystalNucleationLiquids2016}, ionic systems~\cite{bedrovMolecularDynamicsSimulations2019}, and crystalline solids~\cite{zepeda-ruizProbingLimitsMetal2017,cyganAdvancesClayffMolecular2021}, to name a few examples. 
With the rapid adoption of machine-learning interatomic potentials~\cite{behler2007generalized,deringer2019mlpotentials,batzner2022nequip,batatia2022mace} and highly scalable molecular dynamics (MD) codes~\cite{plimpton1995fast,abrahamGROMACSHighPerformance2015,thompson2022lammps}, multi-million-atom simulations are now routine.
In many settings, the dominant bottleneck is no longer generating atomic trajectories but rather extracting reliable structural information from them. 
One example, which is the focus here, is the analysis and identification of crystalline defects, such as vacancies, self-interstitial atoms (SIAs), and grain boundaries (GBs), which are often difficult to differentiate from thermally distorted but otherwise intact lattice motifs.
These defects control much of a material's behavior (strength and plasticity, mass transport, and the accumulation of radiation damage~\cite{zepeda-ruizProbingLimitsMetal2017,nordlund1998defect}), so the fidelity of downstream materials insight rests on identifying them reliably.

At low temperatures (far below the melting point), defect identification is often reliable because local neighborhoods remain close to ideal reference geometries. Classical approaches such as common-neighbor analysis (CNA)~\cite{honeycutt1987cna,faken1994systematic}, centrosymmetry analysis~\cite{kelchner1998centrosymmetry}, Steinhardt-type bond-order parameters~\cite{steinhardt1983bond}, and polyhedral template matching (PTM) can then separate crystal classes with relatively simple thresholds or template comparisons~\cite{stukowski2012structure,larsen2016ptm}. At elevated temperatures, however, thermal motion continuously perturbs atomic positions and erodes class boundaries in descriptor space: the same local environment may shift from ``easy'' to ``ambiguous'' without any topological change in the underlying defect state. Methods that are reliable at low temperatures therefore degrade sharply near melting.

A second challenge is that training data are distributed unevenly across temperatures. Low-temperature structures can be labeled confidently using geometry-based tools~\cite{stukowski2012structure,larsen2016ptm,stukowski2010ovito}, Wigner--Seitz analysis~\cite{nordlund1998defect}, or controlled defect construction. High-temperature structures are precisely where robust analysis is most needed, yet they are the regime in which labels are least trustworthy. 
For example, even if atoms are clearly labeled when an atomic structure is initialized (e.g., constructed by hand), at high temperatures it is commonplace for atoms to rearrange and structures to migrate, thereby rendering the labels unreliable.
This creates a practically important domain-shift problem~\cite{bendavid2010theory}: can structural knowledge learned from well-defined low-temperature configurations be transferred to thermally perturbed configurations without labels at the target temperature?
Additionally, how can the accuracy of the identified structures be tested if no labeled data exist at elevated temperatures?

Two recent lines of work bracket this question without resolving it. Chung \emph{et al.}~\cite{chung2022data} demonstrated that supervised, neural-network-based classifiers trained on synthetically perturbed reference structures outperform geometric heuristics for crystal-structure identification at high temperature; however, their setting is in-sample, with reliable labels assumed at the target temperature, and defect environments are treated as outliers rather than as a class of interest. 
Furthermore, a similar study showed that classification performance with synthetically perturbed liquid structures was quite poor~\cite{erhardCrystalStructureIdentification2024}, a problematic observation given that many defective states are relatively ``disordered'' in comparison to perfect crystal structures. 
Hsu \emph{et al.}~\cite{hsu2024scorebased} introduced a score-based denoiser that suppresses thermal vibrations and allows standard tools such as CNA and PTM to operate near the melting point; in that pipeline, defects appear only as the residual ``other'' category after structure identification, and the analysis remains tied to the template inventory of classical tools. 
Related descriptor- and graph-based classifiers improve robustness within the thermal regime covered by training~\cite{leitherer2021robust,swanson2020deep,takahashi2020mining,becker2022unsupervised}, and learned denoisers have been used for downstream structure recovery more broadly~\cite{dai2024inpainting,zaidi2023pretraining,yang2025generative}. 
In most existing workflows, however, denoising and classification are treated as separate steps, and the central issue of label transfer across temperature is left implicit.

Local structure identification in atomistic simulation has a long history. Widely used methods such as CNA~\cite{honeycutt1987cna,faken1994systematic}, centrosymmetry analysis~\cite{kelchner1998centrosymmetry}, bond-angle analysis~\cite{ackland2006applications}, and PTM compare local neighborhoods to ideal crystal motifs or evaluate symmetry-sensitive descriptors~\cite{stukowski2012structure,larsen2016ptm}. These approaches are computationally efficient and interpretable, but they are inherently sensitive to thermal motion, elastic strain, and local disorder because they depend on fixed tolerances or template similarity. Near melting, the distinction between true defects and temporarily distorted crystal neighborhoods becomes especially unclear.

More robust feature constructions have been proposed, including bond-order spectra~\cite{steinhardt1983bond,lechner2008accurate}, SOAP-like representations~\cite{bartok2013representing}, and supervised combinations of invariant descriptors~\cite{chung2022data,leitherer2021robust}. These improve robustness but remain, in essence, in-sample methods: they work best when the training distribution resembles the inference distribution. The data-centric framework of Chung \emph{et al.}~\cite{chung2022data} is the closest supervised antecedent to our work: it shows that multi-layer perceptron (MLP) classifiers can outperform heuristics at high temperature, but it assumes labeled high-temperature data and does not treat defects as an explicit class. Our framework is, in this sense, its natural cross-temperature, defect-aware extension.
While Chung \emph{et al.} employed an ad hoc set of atomic descriptors, we employ the atomic cluster expansion (ACE) descriptors~\cite{drautz2019ace} because of their completeness property in describing atomic environments and their performance in machine learning interatomic potentials~\cite{batatia2022mace}.

Machine-learning approaches replace hand-tuned thresholds with learned decision boundaries. Descriptor-based classifiers use SOAP~\cite{bartok2013representing}, ACE~\cite{drautz2019ace}, or radial/angular descriptors~\cite{behler2007generalized} as inputs to kernel models, multilayer perceptrons, or tree ensembles; graph neural networks and equivariant architectures learn directly from atomic graphs while preserving Euclidean symmetries~\cite{swanson2020deep,takahashi2020mining,batzner2022nequip,batatia2022mace}. These advances have significantly improved phase and structure classification in noisy settings~\cite{leitherer2021robust,becker2022unsupervised,islam2025dislocation}, and class-incremental formulations further allow the catalog of structure classes to grow over time without retraining on all previously seen classes~\cite{kim2026cil}.

A complementary line targets defects with \emph{unsupervised} learning. Becker \emph{et al.}~\cite{becker2022unsupervised} identify atomic structures without labels via topological learning, and Del~Fr\'e \emph{et al.}~\cite{delfre2025unsupervised} demonstrate an unsupervised SOAP+autoencoder pipeline for defect detection in displacement cascades. K{\'y}vala \emph{et al.}~\cite{kyvala2025unsupervised} likewise identify phases and point defects by unsupervised dimensionality reduction of per-atom descriptors, mitigating thermal noise by time-averaging the descriptors rather than by aligning temperature domains. Such methods are attractive when no labels exist at all, but they do not exploit the rich structural information that \emph{is} available at low temperature. Our setting is different: low-temperature labels exist and are reliable, and the question is how to transfer them to regimes where labeling fails. In supervised settings, most published structure classifiers still assume representative labeled data across the conditions of interest; when labels are trustworthy only at low temperature, naive supervised learning invites overconfident extrapolation, compounded for defects by severe class imbalance.

Score-based and diffusion-inspired models offer a complementary strategy: rather than making the classifier invariant to thermal noise, one first learns a denoising operator that removes perturbations while preserving meaningful structure~\cite{hsu2024scorebased,zaidi2023pretraining}, an approach that extends to structural completion and related generative tasks~\cite{dai2024inpainting,yang2025generative}. Denoisers are naturally trained in a self-supervised manner on synthetically perturbed reference structures~\cite{zaidi2023pretraining}. The most closely related work is the score-based denoising framework of Hsu \emph{et al.}~\cite{hsu2024scorebased}, which showed that iterative denoising dramatically improves downstream CNA/PTM classification near melting and has since been extended to other structure-classification settings such as ice-phase identification~\cite{sun2024ice}. That work provides the input-level component of our pipeline. The point of departure is that, in their setting, defect identification is left to the downstream structure-identification tool, which restricts the analysis to the templates that tool supports and treats defects as the residual ``other'' bin. We instead make defects an explicit learned class and use cross-temperature alignment to transfer reliable low-temperature labels into the regime where labels are least available.
Furthermore, as we shown in Appendix~\ref{appendix:SIA_collapse}, rather than using a pre-trained denoiser, we show that a custom denoiser for each defect class is required to ensure robust denoising without artifacts.



In this work, we make crystalline defects a class of interest in their own right, and pursue a data-driven classification framework which is reliable up to the melting point. 
In this task, three key challenges arise: (1) how to cope with the highly heterogeneous local atomic environments of defects (in comparison to relatively homogeneous crystal structures)? (2) how to train a classifier when reliable labeled data at near-melting temperatures are unavailable? and (3) how to test the performance of the classifier without labeled data?
We demonstrate that a robust classification framework can be built by leveraging score-based denoising~\cite{hsu2024scorebased} as the input-level component, in conjunction with morphological regularization and cross-temperature alignment during training. 
Crucially, we show that denoising alone cannot attain robust performance up to melting. 
Instead, morphological regularization is necessary to discourage fragmented, geometrically implausible predicted defect structures, thereby resolving challenge (1) above. 
To address challenge (2), cross-temperature alignment using \emph{contrastive learning} encourages latent representations of true high-temperature MD data and synthetically perturbed data to be well aligned.
In contrastive learning, a model is trained to pull different but related data (true MD and synthetically noised data in our case) closer together in a latent feature space, while pushing unrelated data further apart~\cite{khosla2020supcon}.
Hence, this scheme involves domain alignment at three distinct levels: \emph{input-level alignment} is achieved through use of score-based denoising, \emph{representation-level alignment} is attained with cross-temperature contrastive learning, and \emph{prediction-level alignment} is enforced through morphological regularization.

Finally, to address challenge (3) relating to testing without labeled data, we have developed a label-free, reference-relative evaluation suite that scores a predicted structure along five complementary axes (spatial coherence, fragmentation, geometric compactness, boundary consistency, and consistency with an independent physics-based reference signal), normalized to a common scale. 
No single axis is decisive; jointly they triangulate what a physically credible atomic structure must look like, enabling model assessment and selection without high-temperature ground truth. 
Where exhaustive ground truth \emph{is} available (for point defects on a known parent lattice, via atom-tracked Wigner--Seitz occupancy analysis), we score the pipeline against it outright, from 4000-atom benchmark cells to a million-atom trajectory, and use the agreement between the two evaluations to demonstrate the merit of the label-free suite where no truth can be constructed.
Relative to prior work, this paper makes five contributions. \textit{\textbf{First}}, we treat defects as an explicit class of interest and reframe high-temperature defect identification as a cross-temperature alignment problem, distinguishing it from in-sample structure classification~\cite{chung2022data} and from denoising-plus-template pipelines in which defects are residuals~\cite{hsu2024scorebased}. \textit{\textbf{Second}}, we combine a learned denoiser (adopted from Ref.~\cite{hsu2024scorebased}, but retrained for each structure of interest), cross-temperature representation alignment, and supervised defect classification into a single end-to-end framework, instead of using denoising as a loosely coupled pre-processing step. \textit{\textbf{Third}}, we introduce a morphology-aware regularization strategy that acts directly on target-domain predictions, encouraging spatially coherent and compact atomic structures while also improving cross-domain alignment through the shared representation; the specific priors used here are one instantiation of a user-customizable family. \textit{\textbf{Fourth}}, we contribute a label-free, reference-relative evaluation suite for predicted defect structures (five complementary spatial and physics-based metrics with a triangulation rationale), applicable whenever target-domain labels are unavailable, and validated here by its agreement with exhaustive Wigner--Seitz ground truth wherever both evaluations can be computed. \textit{\textbf{Fifth}}, we demonstrate the framework across face-centered cubic (FCC), body-centered cubic (BCC), and hexagonal close-packed (HCP) systems containing vacancies and SIAs near melting, scored against Wigner--Seitz ground truth at the defect level; on a million-atom, 2.5\,ns trajectory; and on grain-boundary phase/complexion transformations in aluminum bicrystals, where per-atom classification reveals two distinct nucleation modes. Altogether, these contributions provide significant steps toward a general-purpose atomic-structure classification pipeline.

The Results section is organized around these demonstrations: near-melting point defect identification across six defect--lattice systems, the contribution of each alignment level, ground-truth validation at scale, the extension to extended-defect phases (GBs), and a label-free evaluation that spans them. Details of the learning objectives, datasets, baselines, evaluation metrics, and training protocol are provided in the Methods section.

\section{Results}
\label{sec:experiments}

We evaluate the framework on three parent crystal classes (FCC, BCC, and HCP, using Fe as our element of choice), each in pristine form and containing point defects (vacancies and SIAs), with the central challenge being performance near the melting temperature, where thermal perturbations obscure defect topology in the raw coordinates. The evaluation proceeds in five settings of increasing reach. First, we assess defect identification near melting across all six defect--lattice systems, scored against Wigner--Seitz ground truth at the defect level. Second, we dissect the contribution of each alignment level across temperature under the same ground truth. Third, we deploy the pipeline on a million-atom trajectory, extending the ground-truth validation across three orders of magnitude in scale and time. Fourth, we move beyond point defects to GB phase/complexion transformations, probing generality across defect taxonomy and morphology. Finally, we introduce a label-free evaluation suite, show that it reproduces the ground-truth verdicts where both can be computed, and apply it to the GBs where no ground truth exists.

\subsection{Defect identification near melting}
\label{subsec:qual_results}

\begin{figure}[t]
\centering
\includegraphics[width=\linewidth]{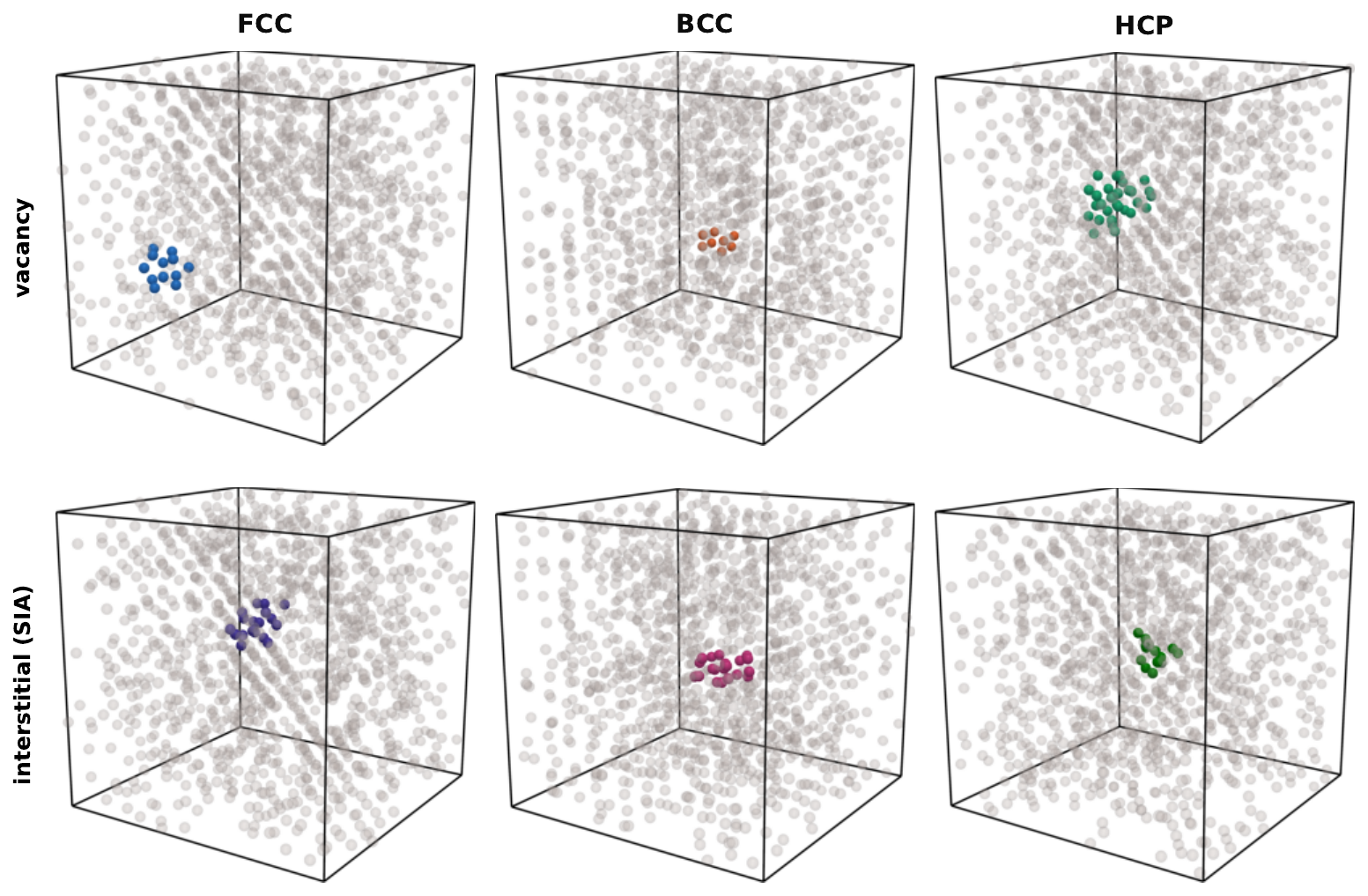}
\caption{Defect detection by the full model (MLP with the contrastive and
morphology terms) across all six systems at the melting temperature. Each panel shows one
representative test snapshot (raw thermal positions); colored atoms are those
the deployed pipeline flags as defective, gray atoms are a subsample of the
remaining bulk. Top row: vacancy-containing systems; bottom row:
self-interstitial-atom-containing systems. In every system the flagged atoms form a
single compact cluster at the defect site, with no scattered false detections
elsewhere in the box; the elongated shape of the interstitial clusters reflects
the dumbbell geometry of the defect. Quantitative per-system results are given in Table~\ref{tab:metrics}.}
\label{fig:qualitative_vac}
\end{figure}

Figure~\ref{fig:qualitative_vac} shows the full model's predictions for representative near-melting snapshots of all six systems: vacancy-containing and SIA-containing FCC, BCC, and HCP crystals. Working directly from raw thermal coordinates, the deployed pipeline flags a single compact cluster of atoms at the true defect site in every system, with no scattered false detections elsewhere in the box. The visual signature of each defect type is preserved: vacancy predictions form roughly isotropic clusters around the missing-atom site, while the elongated SIA clusters reflect the dumbbell geometry of the defect. This is the qualitative behavior that matters for downstream analysis: defect atomic structures that read as localized physical objects rather than diffuse probability clouds.

For the point-defect systems, this qualitative impression can be tested against exhaustive ground truth. Because each system derives from a known parent lattice, atom-tracked Wigner--Seitz occupancy analysis on the denoised coordinates yields a complete registry of true vacancy and interstitial sites even at the melting temperature (Methods), a truth signal used for evaluation only, never for training, so the cross-temperature setting is preserved: the model learns exclusively from low-temperature labels.
Detection is scored at the level of whole defects rather than individual atoms (the \emph{defect-object level}): each predicted defect object (a connected cluster of flagged atoms) either matches a true defect site or it does not, and we report object recall, precision, F1, and the number of false objects per snapshot.

Table~\ref{tab:metrics} reports this evaluation for the deployed pipeline. Three features stand out. First, pristine crystals are read perfectly (Panel~A): no defect objects are reported where none exist, and held-out pristine atoms are classified without error---the pipeline does not hallucinate structure under thermal noise. Second, vacancy detection is essentially solved near melting (Panel~B): perfect object-level detection in FCC and HCP, F1 of 0.90 in BCC, and zero false objects across all three lattices, so every reported vacancy is real; the residual imperfection is a recall dip in BCC, an occasional missed defect rather than an invented one. Third, SIAs show the complementary profile (Panel~C): the pipeline localizes every interstitial (object recall of 1.00 in all three lattices), and the F1 of 0.92 reflects an occasional small spurious cluster away from the defect site, at most a quarter of a false object per snapshot on average. The full model's errors are thus rare and defect-type-specific in a useful way: for vacancies it never fabricates and occasionally misses (BCC), while for SIAs it never misses and occasionally fabricates one small cluster. In both cases the dominant signal---which defects exist and where---is delivered intact near the melting point.

\begin{table}[t]
\centering
\caption{Ground-truth evaluation of the deployed pipeline at the melting temperature. Detection is scored at the defect level against atom-tracked Wigner--Seitz occupancy labels: predicted defect objects (connected clusters of flagged atoms) are matched one-to-one to true defect sites by center-of-mass proximity (Methods), and we report object-level recall, precision, F1, and false objects per snapshot (FP). Entries are mean $\pm$ standard deviation across independent runs; ground truth is used for evaluation only, never for training. Ladder comparisons appear in Table~\ref{tab:ablation}.}
\label{tab:metrics}
\setlength{\tabcolsep}{6pt}
\renewcommand{\arraystretch}{1.15}
\begin{tabular}{@{}l cccc@{}}
\toprule
System & Recall & Precision & F1 & FP \\
\midrule
\multicolumn{5}{@{}l}{\textit{Panel A: Pristine crystals}} \\
\addlinespace[2pt]
Perfect FCC          & $1.00 \pm 0.00$ & $1.00 \pm 0.00$ & $1.00 \pm 0.00$ & $0.00 \pm 0.00$ \\
Perfect BCC          & $1.00 \pm 0.00$ & $1.00 \pm 0.00$ & $1.00 \pm 0.00$ & $0.00 \pm 0.00$ \\
Perfect HCP          & $1.00 \pm 0.00$ & $1.00 \pm 0.00$ & $1.00 \pm 0.00$ & $0.00 \pm 0.00$ \\
\midrule
\multicolumn{5}{@{}l}{\textit{Panel B: Vacancy systems}} \\
\addlinespace[2pt]
FCC + vacancy        & $1.00 \pm 0.00$ & $1.00 \pm 0.00$ & $1.00 \pm 0.00$ & $0.00 \pm 0.00$ \\
BCC + vacancy        & $0.87 \pm 0.27$ & $1.00 \pm 0.00$ & $0.90 \pm 0.20$ & $0.00 \pm 0.00$ \\
HCP + vacancy        & $1.00 \pm 0.00$ & $1.00 \pm 0.00$ & $1.00 \pm 0.00$ & $0.00 \pm 0.00$ \\
\midrule
\multicolumn{5}{@{}l}{\textit{Panel C: Self-interstitial atom systems}} \\
\addlinespace[2pt]
FCC + SIA            & $1.00 \pm 0.00$ & $0.88 \pm 0.22$ & $0.92 \pm 0.14$ & $0.25 \pm 0.43$ \\
BCC + SIA            & $1.00 \pm 0.00$ & $0.88 \pm 0.22$ & $0.92 \pm 0.14$ & $0.25 \pm 0.43$ \\
HCP + SIA            & $1.00 \pm 0.00$ & $0.88 \pm 0.22$ & $0.92 \pm 0.14$ & $0.25 \pm 0.43$ \\
\bottomrule
\end{tabular}
\par\vspace{3pt}
\footnotesize
Pristine systems contain no defects: their rows report that no defect objects are fabricated
and that held-out pristine atoms are classified perfectly; pristine entries are mean $\pm$ s.d.\
across random held-out splits. FP counts false predicted defect objects per snapshot.
\end{table}

\subsection{Temperature robustness and the contribution of each alignment level}
\label{subsec:ablation}

\begin{figure}[htbp]
\centering
\includegraphics[width=\linewidth]{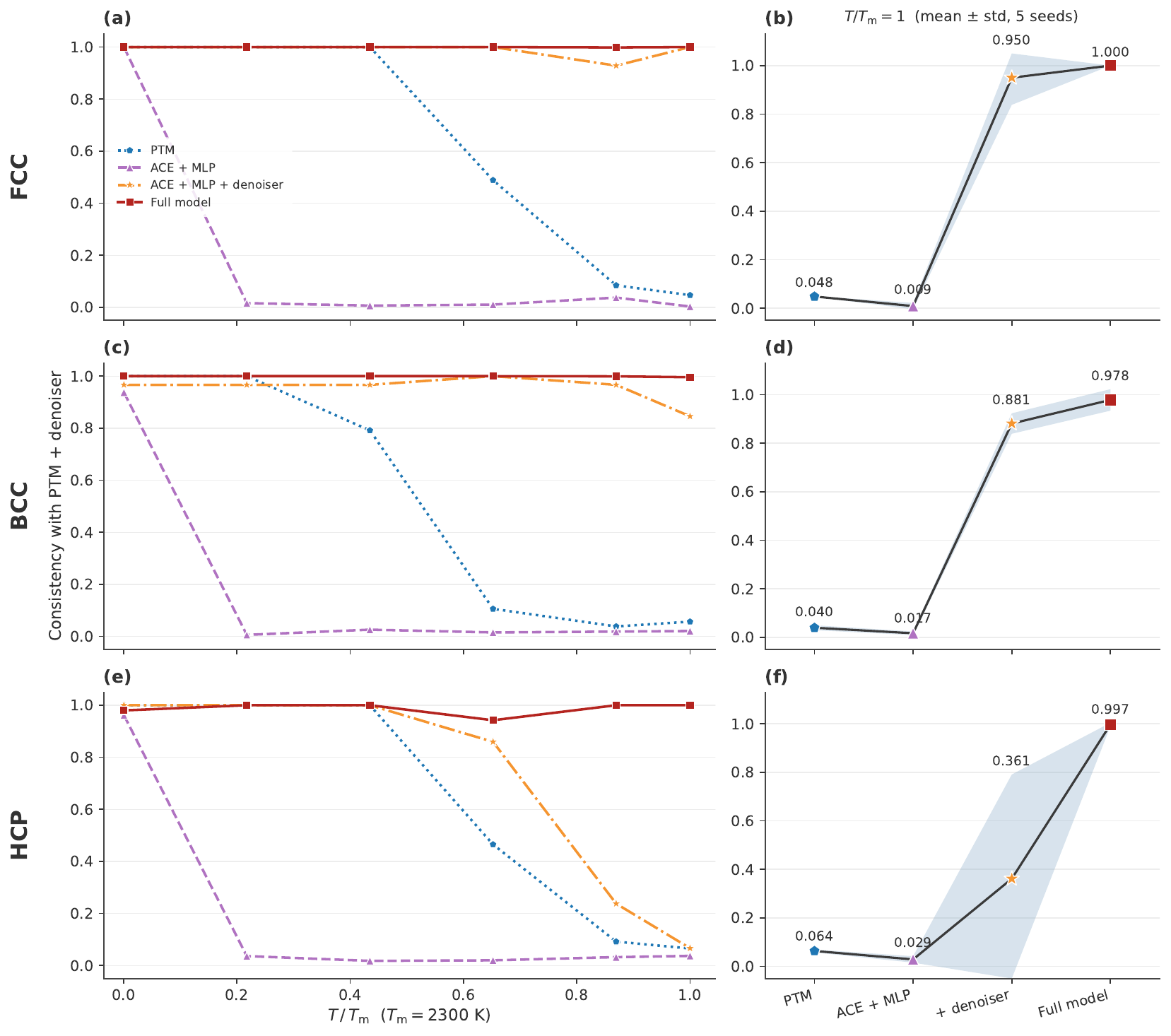}
\caption{Temperature robustness for the vacancy task. \textbf{(a,\,c,\,e)}~Physics-consistency
score (agreement with the denoiser+PTM reference; Methods) as a function of homologous
temperature $T/T_m$ for FCC, BCC, and HCP. \textbf{(b,\,d,\,f)}~The same score at $T/T_m = 1$ for the model ladder,
from the geometric baseline (PTM) through descriptor-based classification (ACE\,+\,MLP),
input-level denoising (+\,denoiser), and the full model. The complete rung-by-rung
dissection, including the intermediate +\,contrastive configuration, is given in
Table~\ref{tab:ablation}. Shaded bands denote one standard deviation across
independent runs. Raw geometric and
descriptor-only classification degrade sharply with temperature; denoising restores most of
the reference consistency in FCC and BCC but much less in HCP, where the full model's
alignment terms are decisive.}
\label{fig:temp_curves}
\end{figure}

Figure~\ref{fig:temp_curves} traces the vacancy task across the full temperature range from 0 K up to melting; (a,c,e) show performance across temperature, while (b,d,f) show performance at the melting temperature with different classifier configurations. At low temperatures all methods are close, because the underlying lattice symmetry remains easy to recognize. As temperature rises, raw geometric classification (PTM~\cite{larsen2016ptm}) and direct descriptor-based classification (ACE\,+\,MLP) collapse; at the melting temperature their physics-consistency scores sit at 0.04--0.06 and 0.01--0.03, respectively, across the three lattices. Input-level denoising~\cite{hsu2024scorebased} restores most of the lost signal in FCC and BCC (0.95 and 0.88 at the melting temperature) but only a fraction of it in HCP (0.36, with large run-to-run spread). Our full model is what closes the gap in all three lattices (1.00, 0.98, 1.00), and it does so with small variance, indicating that the alignment machinery matters most exactly where denoising alone is least sufficient.

Table~\ref{tab:ablation} scores the same ladder against Wigner--Seitz ground truth on the vacancy systems, at the defect-object level and at the melting temperature. The rungs map directly onto the three alignment levels: ``+\,denoiser'' adds input-level alignment, ``+\,contrastive'' adds representation-level alignment, and ``+\,morphology'' adds the prediction-level morphology regularization, yielding the full model. The raw rungs fail absolutely: applied to raw thermal coordinates, neither PTM nor ``ACE+MLP'' detects a single true defect object at the melting temperature---every object they report is false. Input-level denoising turns the task from impossible to merely hard: object recall jumps to 1.00 in FCC and BCC but only 0.40 in HCP, while precision remains as low as 0.22--0.82, with several false objects per snapshot. The remaining levels convert detection into trust. The contrastive term raises precision at held recall in the cubic lattices (FCC reaches perfect detection already at this rung), and the full model finishes the job: object F1 of 1.00/0.90/1.00 for FCC/BCC/HCP with \emph{zero} false objects in every system. The ladder is not uniform on every axis (BCC trades some recall, 1.00\,$\rightarrow$\,0.87, for its perfect precision), but its direction is unmistakable, and on the vacancy task the full model's residual errors are omissions, never fabrications.

\begin{table}[htbp]
\centering
\caption{Ablation study at the melting temperature for vacancy-containing FCC, BCC, and HCP systems, scored against Wigner--Seitz ground truth at the defect-object level (protocol of Table~\ref{tab:metrics}). The rows form a ladder aligned with the three levels of the framework: the geometric baseline (PTM) and descriptor-based classification (ACE\,+\,MLP) operate on raw thermal coordinates; +\,denoiser adds input-level alignment; +\,contrastive adds representation-level alignment; and the full model (+\,morphology) adds the prediction-level morphology term. For recall, precision, and F1, higher is better ($\uparrow$); FP counts false predicted defect objects per snapshot, so lower is better ($\downarrow$). Best results per group are in \textbf{bold}.}
\label{tab:ablation}
\setlength{\tabcolsep}{6pt}
\renewcommand{\arraystretch}{1.15}
\begin{tabular}{@{}llcccc@{}}
\toprule
\textbf{Structure} & \textbf{Model} & \textbf{Recall} $\uparrow$ & \textbf{Precision} $\uparrow$ & \textbf{F1} $\uparrow$ & \textbf{FP} $\downarrow$ \\
\midrule
\multirow{5}{*}{\textbf{FCC}}
  & PTM           & $0.00 \pm 0.00$ & $0.00 \pm 0.00$ & $0.00 \pm 0.00$ & $4.60 \pm 1.02$ \\
  & ACE + MLP     & $0.00 \pm 0.00$ & $0.00 \pm 0.00$ & $0.00 \pm 0.00$ & $7.40 \pm 3.72$ \\
  & + denoiser    & $\mathbf{1.00 \pm 0.00}$ & $0.82 \pm 0.37$ & $0.83 \pm 0.34$ & $2.20 \pm 4.40$ \\
  & + contrastive & $\mathbf{1.00 \pm 0.00}$ & $\mathbf{1.00 \pm 0.00}$ & $\mathbf{1.00 \pm 0.00}$ & $\mathbf{0.00 \pm 0.00}$ \\
  & + morphology   & $\mathbf{1.00 \pm 0.00}$ & $\mathbf{1.00 \pm 0.00}$ & $\mathbf{1.00 \pm 0.00}$ & $\mathbf{0.00 \pm 0.00}$ \\
\midrule
\multirow{5}{*}{\textbf{BCC}}
  & PTM           & $0.00 \pm 0.00$ & $0.00 \pm 0.00$ & $0.00 \pm 0.00$ & $1.00 \pm 0.00$ \\
  & ACE + MLP     & $0.00 \pm 0.00$ & $0.00 \pm 0.00$ & $0.00 \pm 0.00$ & $8.40 \pm 3.83$ \\
  & + denoiser    & $\mathbf{1.00 \pm 0.00}$ & $0.27 \pm 0.12$ & $0.41 \pm 0.14$ & $5.20 \pm 4.21$ \\
  & + contrastive & $\mathbf{1.00 \pm 0.00}$ & $0.55 \pm 0.38$ & $0.63 \pm 0.31$ & $3.40 \pm 3.56$ \\
  & + morphology   & $0.87 \pm 0.27$ & $\mathbf{1.00 \pm 0.00}$ & $\mathbf{0.90 \pm 0.20}$ & $\mathbf{0.00 \pm 0.00}$ \\
\midrule
\multirow{5}{*}{\textbf{HCP}}
  & PTM           & $0.00 \pm 0.00$ & $0.00 \pm 0.00$ & $0.00 \pm 0.00$ & $2.60 \pm 1.85$ \\
  & ACE + MLP     & $0.00 \pm 0.00$ & $0.00 \pm 0.00$ & $0.00 \pm 0.00$ & $2.40 \pm 1.96$ \\
  & + denoiser    & $0.40 \pm 0.49$ & $0.22 \pm 0.39$ & $0.23 \pm 0.39$ & $2.60 \pm 3.72$ \\
  & + contrastive & $0.80 \pm 0.40$ & $0.27 \pm 0.37$ & $0.32 \pm 0.37$ & $35.60 \pm 64.78$ \\
  & + morphology   & $\mathbf{1.00 \pm 0.00}$ & $\mathbf{1.00 \pm 0.00}$ & $\mathbf{1.00 \pm 0.00}$ & $\mathbf{0.00 \pm 0.00}$ \\
\bottomrule
\end{tabular}
\end{table}

\begin{figure}[htbp]
\centering
\includegraphics[width=1.0\linewidth]{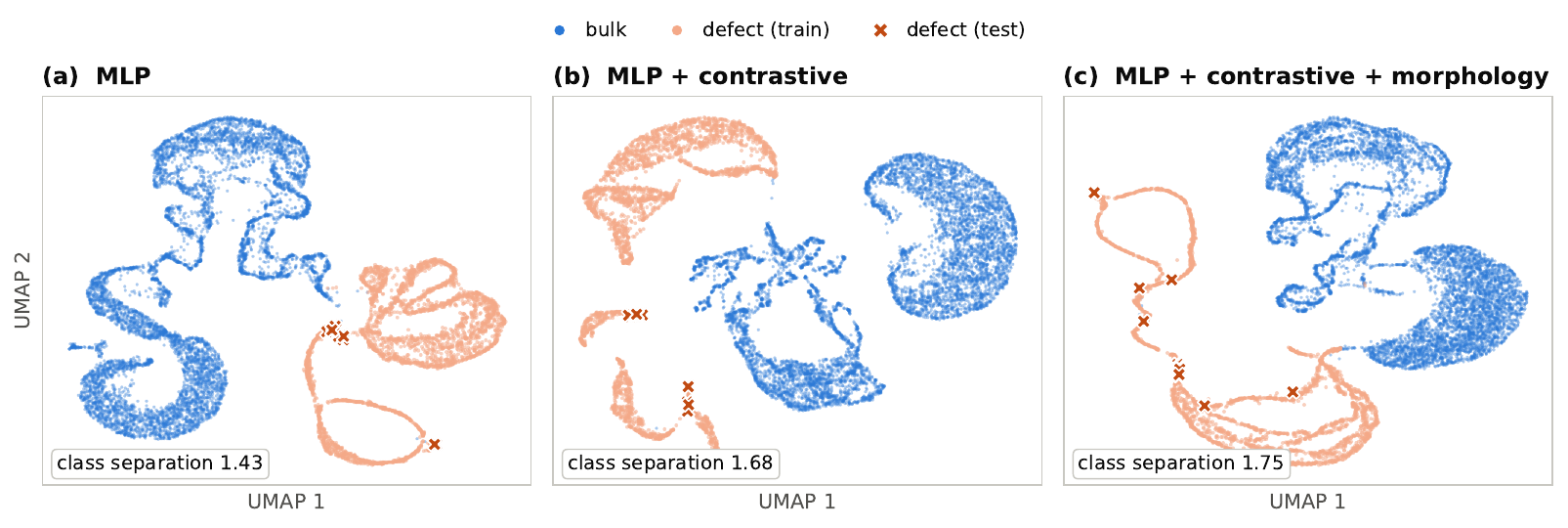}
\caption{Latent-space organization across the model ladder (HCP vacancy system at the melting temperature). Each panel shows a 2-D UMAP embedding of the last-hidden-layer
representations of one model: \textbf{(a)}~plain MLP, \textbf{(b)}~MLP with the
contrastive term, \textbf{(c)}~MLP with the contrastive and morphology terms. All three models are trained on
identical noise-matched training data; only the training objective differs.
All three operate on denoised coordinates, so panels (a)--(c) correspond to
the +\,denoiser, +\,contrastive, and full-model rungs of
Table~\ref{tab:ablation}, respectively.
Blue points are bulk atoms (training and test pooled), orange points are the
defect examples seen during training, and red crosses are the defect atoms of
the unseen MD test snapshot (labels from PTM on the denoised
configuration). Class separation is the distance between the bulk and defect
class centroids in the full latent space after per-dimension standardization
and $L_2$ normalization (range 0--2). Adding the two alignment terms increases
the separation monotonically ($1.43 \rightarrow 1.68 \rightarrow 1.75$): in (a)
the test defects lie in the contact region between the bulk and defect
manifolds, whereas in (b) and (c) the classes form well-separated manifolds and
the unseen test defects fall on the manifold spanned by the training defects,
indicating that the learned separation transfers to the true MD test domain.
The same trend appears in detection performance, where false positives at fixed
recall drop from 3{,}639~(a) to 19~(b) to 0~(c). UMAP
(\texttt{n\_neighbors} = 50, \texttt{min\_dist} = 0.4) is used for
visualization only: it preserves which points are neighbors, but distances in
the 2-D layout carry no quantitative meaning, which is why the class-separation
values quoted above are computed in the full latent space rather than measured
from the plot.}
\label{fig:latent}
\end{figure}

Figure~\ref{fig:latent} makes the representation-level story visible. It embeds the last-hidden-layer representations of the three learned rungs via uniform manifold approximation and projection (UMAP) for an HCP vacancy system at the melting temperature, colored by class. In the plain MLP, the unseen high-temperature test defects fall in the contact region between the bulk and defect manifolds, exactly the ambiguity that in-sample classifiers cannot resolve. Adding the contrastive term separates the manifolds, but the true MD defect data used for testing only span a small subset of the manifold occupied by the synthetic training data.
Finally, adding the morphology term separates them further (the class-centroid separation increases monotonically as 1.43\,$\rightarrow$\,1.68\,$\rightarrow$\,1.75) and the MD test defects land on the manifold spanned by the synthetic training defects, indicating that the learned separation transfers to the thermal test domain. The same trend appears in detection performance, where false positives at fixed recall drop from 3{,}639 to 19 to 0 across the three panels. This is direct evidence for a claim that matters beyond this task: a prediction-level prior, optimized end-to-end, improves the learned representation itself: the morphology term is defined on model output, yet its effect is visible in latent space.

\subsection{Ground-truth validation in a million-atom simulation}
\label{subsec:bigsim}

The benchmark cells above certify the pipeline against ground truth in small, controlled systems. The practical question is whether that fidelity survives the regime the framework was actually built for: simulations at the scale where no human or heuristic labeling is feasible. We therefore deployed the pipeline on a one-million-atom FCC Fe simulation at $T/T_m = 0.91$ under 3\,GPa hydrostatic tension, containing on the order of a thousand vacancies and a thousand SIAs, with frames spanning 10\,ps to 2.49\,ns of annealing. Here, as in the benchmark cells, atom-tracked Wigner--Seitz occupancy analysis on the denoised coordinates provides per-site ground truth against which every prediction can be scored.

\begin{figure}[htbp]
\centering
\includegraphics[width=\linewidth]{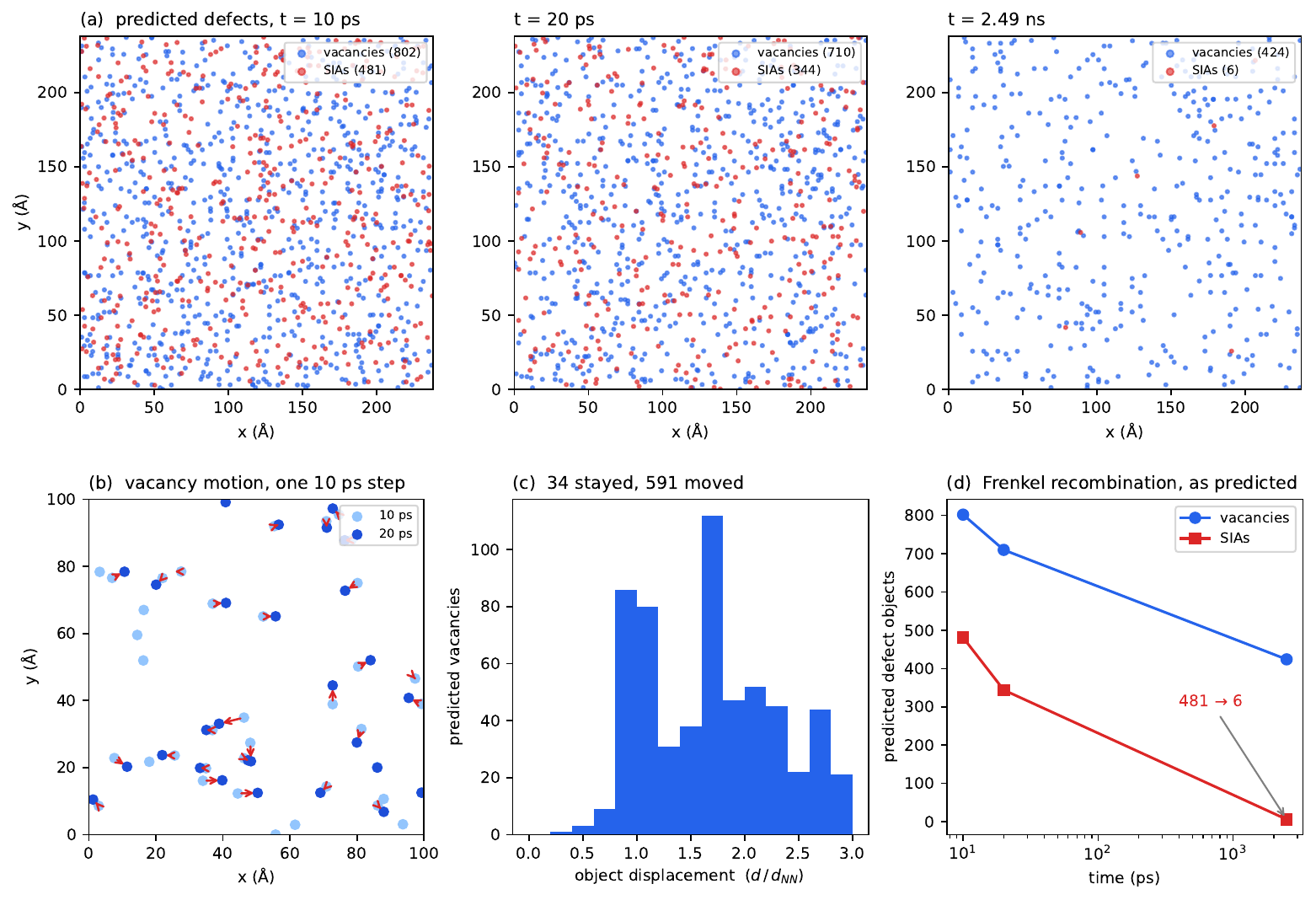}
\caption{\textbf{The classifier tracks point defects through 2.5\,ns of
annealing with no retraining.} All panels show classifier predictions only; no
ground truth enters the figure.
\textbf{(a)}~Predicted defect objects (clusters of predicted atoms, shown as
one point per defect) in the 1M-atom FCC Fe frame at $t = 10$\,ps, 20\,ps,
and 2.49\,ns. The predicted counts (802/710/424 vacancies, 481/344/6 SIAs)
track the Wigner--Seitz truth (796/707/423 and 433/307/0 defect groups) to
within 1\% for vacancies at every time.
\textbf{(b)}~Predicted vacancies in a $100 \times 100 \times 60$\,\AA{} slab
at 10\,ps (light) and 20\,ps (dark); arrows connect each vacancy to its
matched position one frame later, resolving individual lattice hops.
\textbf{(c)}~Displacement distribution of the matched vacancies (34
unchanged, 591 moved within $3\,d_\mathrm{NN}$).
\textbf{(d)}~Predicted defect-object counts versus time: vacancies and
SIAs disappear in near-equal numbers, and by 2.49\,ns the model
reports an SIA-free crystal ($481 \rightarrow 6$ residual objects,
36 atoms out of $10^{6}$), reproducing the complete Frenkel-pair
recombination that the ground truth confirms ($1054 - 599 = 455$ vs.\ 453
surviving vacancies). Scored against the Wigner--Seitz truth
($2.0\,d_\mathrm{NN}$ core+halo), the pipeline finds 1054/1054
vacancies and 597/599 interstitials at 10\,ps, with 15 residual false
vacancy atoms per million.}
\label{fig:big_sim}
\end{figure}

Figure~\ref{fig:big_sim} summarizes the outcome. At the defect level (clustering defective atoms into discrete vacancies and SIAs), the pipeline finds 1054 of 1054 vacancies and 597 of 599 interstitials in the first frame, with 15 residual false vacancy atoms per million. Just as important is what happens over time: with no retraining of any kind, the predicted defect populations track the ground truth across a 250-fold span in time (predicted vacancy counts stay within 1\% of truth at every analyzed frame), and the predictions are sharp enough to resolve individual defect migration events, matching predicted vacancies frame-to-frame to reveal single lattice hops. At the endpoint, the model reports an SIA-free crystal (481\,$\rightarrow$\,6 residual objects, 36 atoms out of $10^6$), reproducing the complete Frenkel-pair recombination that the ground truth confirms ($1054 - 599 = 455$ vs.\ 453 surviving vacancies). The physics of the trajectory (vacancy--SIA recombination proceeding to completion, with the surviving vacancy population conserved) is readable directly from the predictions.

This experiment extends the ground-truth story of Tables~\ref{tab:metrics} and~\ref{tab:ablation} across three orders of magnitude in system size and two and a half in time: the defect-level fidelity established in 4000-atom benchmark cells persists in million-atom MD over nanoseconds, with no retraining and no per-frame tuning.

\subsection{Beyond point defects: grain-boundary phase transformations}
\label{subsec:gb}

Point defects are a difficult case for spatial priors (sparse, small, and locally ambiguous), but they are not the only defect class of scientific interest. As a test of generality across defect taxonomy and morphology, we applied the pipeline to a qualitatively different problem: classifying the internal structural phase of GB atoms in Al bicrystals along MD trajectories. For each of two GBs, two competing GB phases/complexions~\cite{frolov2013structural,cantwell2014complexions,zhu2018predicting,meiners2020observations,langenohl2022dual,brink2023universality,choi2025faceting} are defined via minimized reference structures; trajectory frames are denoised, per-atom ACE descriptors are computed, and boundary atoms are classified between the two phases/complexions. 
Each of these GB phases/complexions has a distinctive local atomic structure, which we term ``diamond,'' ``ladder,'' ``kite,'' and ``L'' below.
As in the point-defect setting, labels exist only for the clean reference states, and the classifier must operate on thermally perturbed configurations it has never seen: the same cross-temperature label-transfer problem, now for an extended defect.

\begin{figure}[htbp]
\centering
\includegraphics[width=\linewidth,height=0.58\textheight,keepaspectratio]{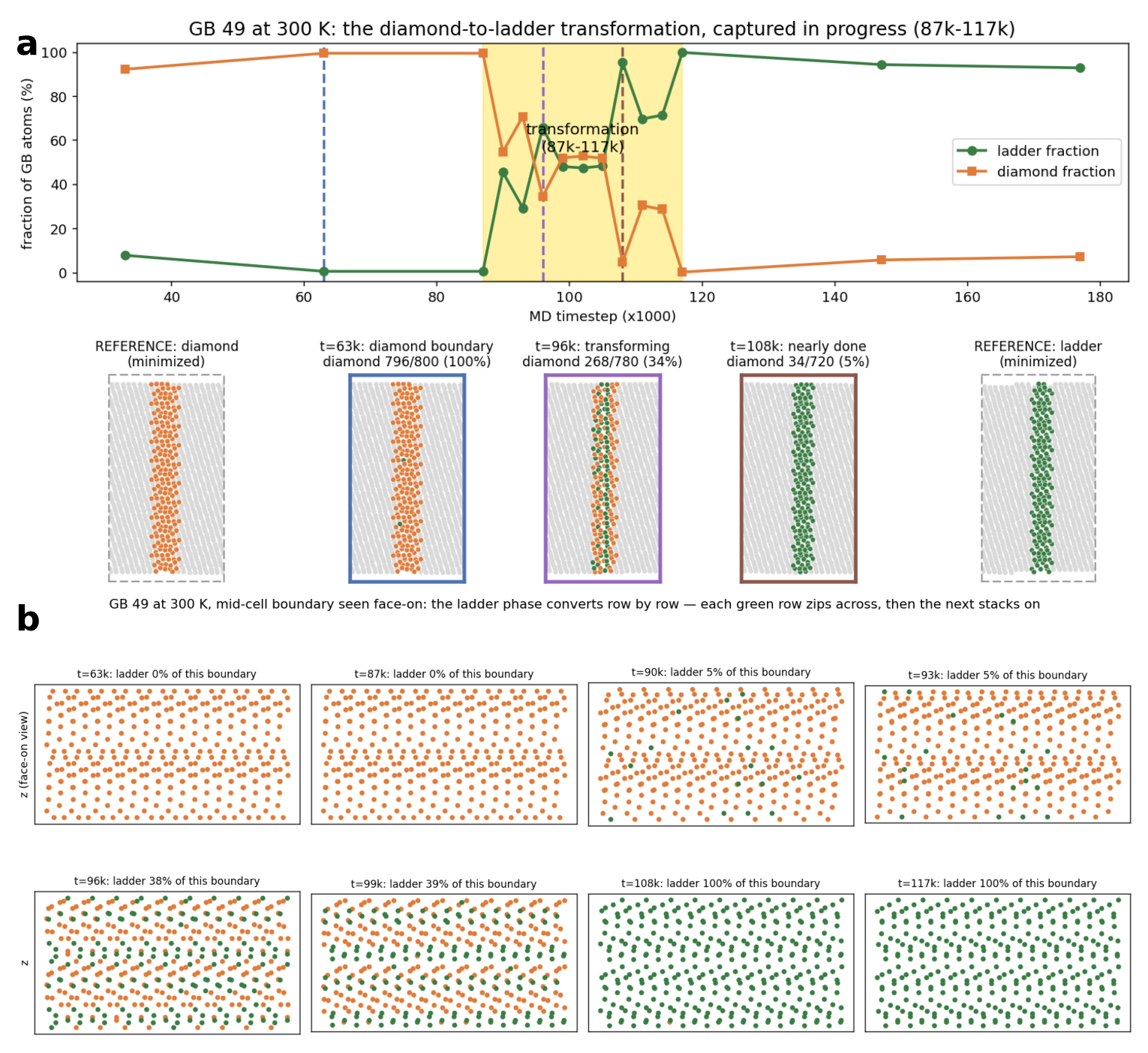}
\caption{A grain-boundary phase transformation captured in progress ($\Sigma11(2\,1\,1)$, $T/T_m=0.29$). Per-atom classification of boundary structure along the MD trajectory: frames are denoised, per-atom ACE descriptors computed, and boundary atoms classified by a classifier trained on the two minimized reference phases (orange = diamond, green = ladder). (a) Fractions of boundary atoms in each phase versus timestep. The diamond$\rightarrow$ladder transformation proceeds gradually over steps 87k–117k (shaded), with transient partial reversals. Below: edge-on maps at representative times, flanked by the reference structures (dashed outlines); counts and fractions above each map, with each trajectory map outlined in the color of its dashed time marker in the curve. (b) The mid-cell boundary viewed face-on (all its atoms in the y–z plane): conversion proceeds row by row — each new ladder row runs nearly the full width of the boundary within one frame interval, rows stack one after another, and the in-fill between them completes the plane. The simulation cell's second boundary converts by the same mechanism, earlier; both show transient near-total retreats before locking in at 117k. Classifier validation: 0.944 leave-one-realization-out; clean references read at 99.8\%/0.1\%; results unchanged under an independently trained denoiser.}
\label{fig:GB49}
\end{figure}

\begin{figure}[htbp]
\centering
\includegraphics[width=\linewidth,height=0.62\textheight,keepaspectratio]{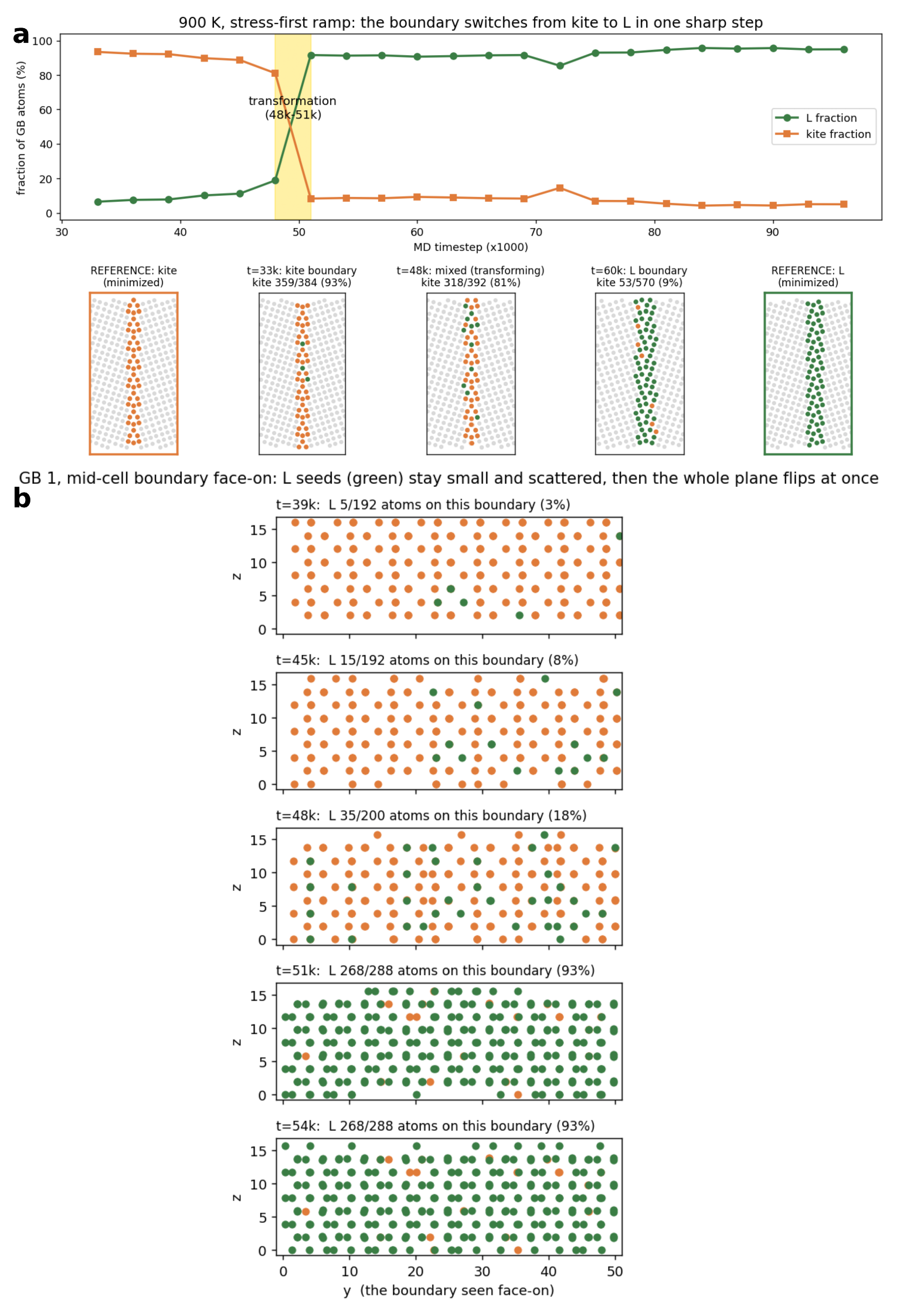}
\caption{GB phase/complexion transformation in a different mode: avalanche conversion under load ($\Sigma5(3\,1\,0)$ symmetric tilt, $T/T_m=0.87$). Layout and color convention as in Fig.~\ref{fig:GB49} (orange = kite, green = L). (a) The kite→L transformation completes in a single sharp step between steps 48k and 51k (shaded). The curve ends before the high-temperature regime where the boundary thickens and per-atom labels become unreliable. (b) Face-on view of the mid-cell boundary: L-classified atoms remain small, scattered seeds through 48k (largest connected cluster \(\leq\) 12 atoms at 18\% converted), then the entire plane converts within one frame interval — a collective avalanche, in contrast to the sequential row-by-row conversion of Fig.~\ref{fig:GB49}b. The boundary thickens upon transforming (192 → 288 atoms on this plane). Classifier validation: 0.996 leave-one-snapshot-out; clean references read at ~100\%/~0\%.}
\label{fig:GB1}
\end{figure}

The per-atom phase maps capture two grain-boundary phase transformations in progress, in two distinct dynamical modes. In the first boundary (Fig.~\ref{fig:GB49}), held at $T/T_m=0.29$ (this temperature was chosen based on when the GB phase/complexion transformation was observed) with no applied load, the diamond-to-ladder transformation proceeds gradually over a 30{,}000-step window: viewed face-on, conversion advances row by row, each new ladder row running nearly the full width of the boundary within one frame interval and successive rows stacking until in-fill completes the plane: sequential, row-by-row nucleation and growth~\cite{winter2022nucleation}. Both boundaries in the cell transform by this mechanism, each with transient near-total retreats before the ladder phase locks in. In the second GB (Fig.~\ref{fig:GB1}), driven at $T/T_m=0.87$ after a stress ramp, the kite-to-L transformation instead completes in a single sharp step: L-classified atoms remain small, scattered seeds for tens of thousands of steps, then the entire boundary plane converts within one frame interval, a collective avalanche, accompanied by a thickening of the boundary as a result of L formation. Distinguishing these two nucleation modes requires exactly what per-atom classification provides: spatially resolved phase assignments over time, not just aggregate order parameters.

The instruments behind these maps are validated in the regime where checks are possible: grouped cross-validation on the reference structures yields accuracies of 0.944 and 0.996 for the two boundaries, the clean references are read essentially perfectly, and the trajectory-level conclusions are unchanged under an independently trained denoiser. Together with the point-defect results, this study supports the framework's central design premise: the components are modular, and the same denoise--describe--classify core, configured per task, extends from point defects in Fe to GB phases/complexions in Al.

\subsection{A label-free evaluation that spans defect types}
\label{subsec:closeness}

Ground truth of the kind used so far exists only because point defects live on a known site registry: Wigner--Seitz occupancy is defined by the parent lattice. Most defect analysis is not so fortunate. For the GB phases/complexions of Section~\ref{subsec:gb} there is no occupancy-style truth at temperature, and the same holds for extended defects generally. For exactly this regime we introduce a label-free evaluation suite, in the spirit of unsupervised segmentation evaluation~\cite{zhang2008survey}, internal cluster-validity indices~\cite{rousseeuw1987silhouettes,davies1979cluster}, and no-reference image-quality assessment~\cite{mittal2013niqe}: five atomic structural metrics (coherence, fragmentation, compact geometry, boundary consistency, and physics consistency), each reported as a closeness, $1-\lvert m_{\mathrm{model}}-m_{\mathrm{ref}}\rvert$, between the metric evaluated on the predicted defect structure and on an independent, physics-based reference (denoiser-assisted PTM), so that 1 indicates exact agreement with reference behavior (Methods; Appendix~\ref{appendix:metrics}). No single axis is decisive; jointly they triangulate what a physically credible defect structure must look like, and a random-permutation baseline anchors the scale from below.

Table~\ref{tab:defect_field_closeness} applies the suite to every defect type in this paper. On the point-defect systems (Panels~A and~B), where the preceding sections provide ground truth, the label-free scores reproduce the truth's verdicts: the vacancy systems that are perfect under Wigner--Seitz scoring sit at closeness 1.00, the one vacancy system with a ground-truth blemish (BCC) is exactly the one the suite flags, and the SIA panel (the defect type that retains residual false detections under ground truth) sits systematically below the vacancy panel. This agreement supports extending the suite to the cases where truth is absent. Panel~C extends it to the GB systems at their respective temperatures: the boundary-phase fields score 0.98--1.00 on every axis, against random baselines as low as 0.02 on physics consistency. The suite thus remains informative for an extended, planar defect class that no occupancy analysis can score---precisely the regime it was designed for.

\begin{table}[t]
\centering
\caption{Defect structure fidelity across defect types. Each entry is the closeness
$1-\lvert m_{\mathrm{model}}-m_{\mathrm{ref}}\rvert$ between a defect structure metric
evaluated on the predicted structure and on the PTM + denoiser reference, for coherence
(Coh.), fragmentation (Frag.), compact geometry (Comp.), boundary (Bound.), and
physics consistency (Phys.); higher is better, with 1 indicating exact agreement
with the reference. Panels A and B use the full model, trained at
$T=0$ and evaluated at the melting temperature; entries are means over independent MD seeds.
Panel C evaluates the pipeline on the two bicrystals of Section~\ref{subsec:gb}, GB1
(Fig.~\ref{fig:GB1}) and GB49 (Fig.~\ref{fig:GB49}), scoring the two competing phases
of each boundary separately at its respective temperature, one trajectory each. \textit{Random} rows report the same closeness
for random permutations of each model's per-atom defect probabilities (25
permutations, pooled within each panel). For the extended planar defects of
Panel C, Comp.\ measures agreement of the defect-region shape statistics with the
reference rather than compactness per se.}
\label{tab:defect_field_closeness}
\setlength{\tabcolsep}{6pt}
\renewcommand{\arraystretch}{1.15}
\begin{tabular}{@{}l ccccc@{}}
\toprule
 & \multicolumn{5}{c}{\textbf{Defect-structure metrics}} \\
\cmidrule(lr){2-6}
System & Coh. & Frag. & Comp. & Bound. & Phys. \\
\midrule
\multicolumn{6}{@{}l}{\textit{Panel A: Vacancy systems (cross-temperature transfer)}} \\
\addlinespace[2pt]
FCC + vacancy               & 1.00 & 1.00 & 1.00 & 1.00 & 1.00 \\
BCC + vacancy               & 0.90 & 0.87 & 0.97 & 0.79 & 0.98 \\
HCP + vacancy               & 1.00 & 0.99 & 1.00 & 1.00 & 1.00 \\
\addlinespace[1pt]
\quad\textit{Random} & \textit{0.78} & \textit{0.09} & \textit{0.81} & \textit{0.57} & \textit{0.06} \\
\midrule
\multicolumn{6}{@{}l}{\textit{Panel B: Interstitial (SIA) systems (cross-temperature transfer)}} \\
\addlinespace[2pt]
FCC + SIA                   & 0.97 & 0.94 & 0.88 & 0.94 & 0.83 \\
BCC + SIA                   & 0.95 & 0.93 & 0.82 & 0.90 & 0.82 \\
HCP + SIA                   & 0.89 & 0.73 & 0.86 & 0.83 & 0.75 \\
\addlinespace[1pt]
\quad\textit{Random} & \textit{0.79} & \textit{0.05} & \textit{0.86} & \textit{0.57} & \textit{0.05} \\
\midrule
\multicolumn{6}{@{}l}{\textit{Panel C: Grain-boundary systems}} \\
\addlinespace[2pt]
GB1 kite ($T/T_m=0.87$)       & 0.99 & 0.99 & 0.99 & 0.98 & 1.00 \\
GB1 L ($T/T_m=0.87$)          & 0.99 & 1.00 & 1.00 & 0.98 & 1.00 \\
GB49 diamond ($T/T_m=0.29$)   & 1.00 & 1.00 & 1.00 & 1.00 & 1.00 \\
GB49 ladder ($T/T_m=0.29$)    & 1.00 & 1.00 & 1.00 & 1.00 & 1.00 \\
\addlinespace[1pt]
\quad\textit{Random} & \textit{0.81} & \textit{0.64} & \textit{0.79} & \textit{0.68} & \textit{0.02} \\
\bottomrule
\end{tabular}
\end{table}

\section{Discussion}
\label{sec:discussion}

The pipeline aligns the low- and high-temperature domains at three coupled levels (input, representation, and prediction), and the results in Section~\ref{subsec:ablation} show all three contributing. The more interesting questions are \emph{why} this particular decomposition works, what the label-free evaluation buys, and when the same recipe should transfer to other atomistic domain shifts.

\paragraph{Why the three levels are complementary rather than redundant.}
The three components are not interchangeable instruments aimed at the same gap; they target different residuals. The denoiser closes the \emph{marginal} mismatch in $p(x)$: thermal noise is the physical cause of the cross-temperature shift, and removing it at the input attacks the largest, most structured component of the gap before any learning takes place. However, denoising is iterative and stochastic, and as Fig.~\ref{fig:latent} shows, it leaves a residual discrepancy concentrated where the denoiser's prior is weakest: near defect cores, where local geometry departs most from the bulk manifold. This residual is precisely a \emph{conditional} ambiguity: after denoising, the same descriptor can still correspond to a distorted bulk environment at high temperature or a defect-affected environment at low temperature. Contrastive learning addresses it because identity-preserved atom pairs supply ground-truth conditional matching without target labels: the $i$-th atom in the low-temperature snapshot and the $i$-th atom in its high-temperature continuation are, by construction, the same physical site, so pulling their embeddings together constrains exactly the ambiguity that survives denoising. Finally, the morphology term addresses the failure mode of the pairing itself: the atoms for which correspondence is least reliable (those near defect cores that have moved or reorganized) are exactly the atoms whose \emph{neighbors} still carry corrective spatial evidence, which is what a graph-based morphological prior exploits. The ablations reflect this division of labor (Table~\ref{tab:ablation}): without denoising, nothing true is detected at all; denoising restores detection but not trust; the contrastive term tightens precision; and the morphology term eliminates fabricated objects outright, most visibly in HCP, where no proper subset of the stack suffices.

\paragraph{What the label-free evaluation buys.}
A methodological contribution of this work is an evaluation that does not stop where labels stop. Where truth can be constructed, we use it outright: Tables~\ref{tab:metrics} and~\ref{tab:ablation} and the million-atom study are scored directly against Wigner--Seitz ground truth. The label-free closeness suite exists for everything beyond that boundary, and its credibility rests on the agreement demonstrated in Section~\ref{subsec:closeness}.
The suite is also honest about its construction: closeness is measured to a physics-based reference pipeline, so a model that mimics that reference scores well by design.
Comparison with a random-permutation baseline provides one safeguard, anchoring every panel from below; the external check against ground truth, wherever one exists, provides the other. We expect this pattern (truth where truth exists, triangulated closeness where it does not, and the two tied together wherever both apply) to be reusable in any dense-prediction problem on atomistic data where labels run out.

\paragraph{When the recipe should transfer beyond cross-temperature shift.}
The framework is presented for cross-temperature transfer, but the underlying decomposition (marginal alignment via a domain-appropriate denoiser, conditional alignment via identity-preserved pairs, prediction-level alignment via a morphological prior) is not specific to temperature. The recipe applies whenever the source-to-target shift admits a learned input-space correction whose residual is local and geometrically structured; this is plausibly true for pressure shifts, modest chemical disorder in dilute alloys, and time-evolution shifts within a single trajectory. The contrastive component specifically requires identity-preserved positive pairs or a substitute. Cross-temperature MD provides pairs naturally, because the high-temperature configuration is the thermal evolution of the low-temperature one; where correspondence breaks (different chemistries, independently generated cells, large-scale diffusion), the contrastive term can be replaced by distributional alternatives such as MMD~\cite{long2015dan}, CORAL~\cite{sun2016deepcoral}, or adversarial alignment~\cite{ganin2015dann}. The other two levels do not depend on pairing and should transfer without modification, consistent with the temperature curves, where denoising alone already recovers most of the reference consistency in FCC and BCC (Fig.~\ref{fig:temp_curves}).

\paragraph{Implications for defect analysis as a learning problem.}
A broader implication is that defect identification is not a special case of phase classification and should not be modeled as one. Defect environments are sparse, structurally diverse, and ambiguous under thermal fluctuation; treating them as a residual ``other'' bin discards exactly the structure that makes them learnable. Treating them as explicit classes with cross-temperature supervision produces representations that are simultaneously more discriminative (clearer class clustering in Fig.~\ref{fig:latent}) and better transferred across domains (test defects landing on the training-defect manifold), a nontrivial combination, since domain-adaptation methods typically trade one for the other. The mechanism, we suspect, is that prediction-level priors on target data act as a supervision channel that the source loss cannot provide, delivered in a form (spatial coherence) that the encoder can internalize. The GB study adds a complementary lesson: once defects are explicitly represented in the label space, ``defect identification'' naturally generalizes from \emph{detecting} defects to \emph{classifying defect-internal structure}, and the same machinery that separates vacancies from bulk separates one GB phase/complexion from another.

We presented a unified framework for cross-temperature atomic structure and defect identification. The method treats defects as classes of interest in their own right and aligns the labeled low-temperature and unlabeled high-temperature domains at three levels: input-level alignment through equivariant denoising (adopted from~\cite{hsu2024scorebased}), representation-level alignment through cross-temperature contrastive learning, and prediction-level alignment through morphology-aware regularization, with the latter also improving the learned representation through end-to-end optimization. Alongside the predictive framework, we introduced a label-free, reference-relative evaluation suite that makes principled model assessment possible in exactly the regime where labels do not exist.

Across vacancy- and SIA-containing FCC, BCC, and HCP systems, the pipeline yields stable, compact, and interpretable defect predictions near melting without labeled high-temperature data. Deployed at scale, it reproduces atom-tracked Wigner--Seitz ground truth in a million-atom, 2.5\,ns trajectory (down to individual vacancy hops and complete Frenkel-pair recombination), anchoring the label-free evaluation to absolute truth. Applied beyond point defects, the same modular core captures grain-boundary phase transformations in aluminum bicrystals and distinguishes two nucleation modes. Just as importantly, the paper proposes a clean conceptual framing: temperature-robust defect analysis is a multi-level domain-alignment problem~\cite{bendavid2010theory,ganin2015dann}, and aligning the domains in input, representation, and prediction space provides both a practical solution here and a template for future work on extended defects, alloy chemistries, and on-the-fly analysis of large-scale simulations.

\subsection{Limitations and outlook}
\label{subsec:limitations}

The framework rests on assumptions and scope choices: some fundamental to the cross-temperature setting, others deliberate choices for this study. We discuss each and indicate where it can be relaxed.

\paragraph{Atom-level correspondence assumption.} The contrastive loss is \emph{atom-wise}: it pairs the $i$-th atom of a low-temperature snapshot with the $i$-th atom of a high-temperature snapshot of the same system, exploiting the fact that the latter is the thermal evolution of the former. This is the natural setting for many MD analysis workflows, but it does not apply universally. When atoms diffuse (most importantly when vacancies migrate), identity-based correspondence becomes unreliable near defect cores; we partially absorb this by reliability-weighting each pair (Eq.~\eqref{eq:weighted_ctr}), so that pairs whose embeddings have already drifted apart contribute little, and the encoder learns from the subset of pairs that still carry signal. Distributional substitutes for broken correspondence are discussed above.

\paragraph{Defect taxonomy.} The classification ladder studied quantitatively here covers point defects in single-element FCC, BCC, and HCP systems, and the morphology priors used in training favor compact, weakly fragmented regions, appropriate for point defects, but not as-is for extended one-dimensional defects (dislocation lines) or two-dimensional defects (stacking faults)~\cite{islam2025dislocation}. The GB study shows that the framework's core extends to interface phases when configured for the task, but a systematic treatment of extended defects using the line- and surface-shaped priors already implemented remains the most immediate direction for follow-up work.

\paragraph{Evaluation.} Wigner--Seitz defect-object ground truth anchors the point-defect results (Tables~\ref{tab:metrics} and~\ref{tab:ablation}, Section~\ref{subsec:bigsim}), but it is tied to a known parent-lattice site registry. Beyond that regime, evaluation is reference-relative: the closeness suite reports agreement with a strong physics-based reference rather than accuracy in the strict sense, and its verdicts are validated by consistency with ground truth only where both can be computed (Section~\ref{subsec:closeness}). Broadening the set of ground-truth-anchored benchmarks (for example, constructed interface configurations with known phase labels) is a worthwhile direction.

\paragraph{Source diversity.} Performance depends on the diversity of the low-temperature reference data and on the extent to which synthetic perturbations span the relevant thermal distortions. Stress, chemical disorder, and finite-size effects are not yet systematically explored.

Beyond these, three extensions appear most promising: generalizing to alloys and multicomponent systems, where chemical and topological disorder interact~\cite{sheriff2024quantifying}; uncertainty quantification near the bulk--defect decision boundary; and integration into on-the-fly MD analysis, for which the pipeline's modular, per-snapshot structure is well suited.

\section{Methods}
\label{sec:methods}

\subsection{Problem formulation}
\label{subsec:problem_setup}

Let $x \in \mathcal{X}$ denote a local atomic environment centered at atom $i$,
\begin{equation}
    x = \mathcal{N}_i = \{(\mathbf{r}_{ij}, s_j): \|\mathbf{r}_{ij}\| \le r_c\},
\end{equation}
where $\mathbf{r}_{ij}=\mathbf{r}_j-\mathbf{r}_i$ are relative coordinates, $s_j$ are species labels, and $r_c$ is the local cutoff radius. Let $y \in \mathcal{Y}$ denote the local structure class. In the systems studied here, $\mathcal{Y}$ includes ideal crystal classes and explicit defect classes such as vacancy-affected or SIA-affected environments.

We assume access to a labeled source dataset of low-temperature environments,
\begin{equation}
    \mathcal{D}_s = \{(x_k,y_k)\}_{k=1}^{N_s}, \qquad x_k \sim p_s(x),
\end{equation}
and an unlabeled target dataset of high-temperature environments,
\begin{equation}
    \mathcal{D}_t = \{x_\ell\}_{\ell=1}^{N_t}, \qquad x_\ell \sim p_t(x).
\end{equation}
The learning problem is to build a predictor that remains reliable on $p_t$ without labels drawn from $p_t$. This is a cross-temperature instance of unsupervised domain adaptation~\cite{bendavid2010theory,ganin2015dann}, in which the source--target shift is induced by thermal perturbation rather than by a change of system or chemistry. The conditional $p(y\mid x)$ is essentially preserved across temperatures (a vacancy is a vacancy in either regime); the marginal $p(x)$ shifts because thermal motion redistributes local environments in descriptor space. The denoiser is designed to reduce this marginal shift; the contrastive and morphology terms close the residual gap.

\subsection{Framework overview}
\label{subsec:pipeline}

\begin{figure}[t]
\centering
\includegraphics[width=\linewidth]{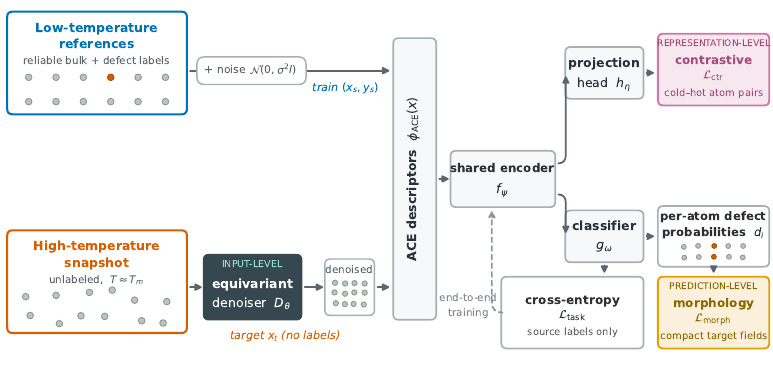}
\caption{Overview of the framework. A labeled source domain (low-temperature
references with bulk and defect classes, broadened by synthetic perturbations)
and an unlabeled target domain (high-temperature snapshots) are aligned at
three levels: an equivariant denoiser aligns the domains in input space, a
contrastive objective on matched cold--hot atom pairs aligns their latent
representations, and a morphology regularizer aligns target-domain predictions
with the compact, coherent geometry of physical defect fields. All terms are
optimized end-to-end through the shared encoder.}
\label{fig:pipeline}
\end{figure}

Figure~\ref{fig:pipeline} summarizes the workflow. From low-temperature (e.g., 0 K) reference structures we construct a labeled training set of local environments containing both bulk-crystal and defect classes, broadening the support of the source distribution with synthetic Gaussian perturbations during training. At inference, each high-temperature snapshot is denoised by an equivariant denoiser~\cite{hsu2024scorebased,batzner2022nequip}, converted into ACE descriptors~\cite{drautz2019ace}, and passed through an MLP classifier that outputs per-atom class probabilities. During training, the classifier is additionally optimized with a cross-temperature contrastive term and a target-domain morphology term; at inference, a light spatial refinement step may optionally be applied to suppress isolated spurious detections. The three learned components correspond to the three alignment levels of the framework, acting in input space, representation space, and output space respectively, with the output-level term also feeding back into the representation through end-to-end optimization. The components are modular: each can be configured, reweighted, or substituted to match the defect class and analysis task at hand, which is how the same core is deployed across the point-defect, large-scale, and grain-boundary studies in this paper.

\subsection{Reference structures and label construction}
\label{subsec:data_generation}

For each parent lattice (FCC, BCC, and HCP), we consider both pristine crystals and configurations containing point defects. 
For the GB cases, reference configurations are obtained via energy minimization, possibly after quenching from elevated temperature if the structure of interest is metastable.
Structural labels are defined using low-temperature reference configurations, where thermal ambiguity is minimal and local environments can be identified reliably with established geometric tools~\cite{stukowski2012structure,larsen2016ptm,nordlund1998defect}. Defect classes (vacancy-affected and SIA-affected environments) are generated either by controlled defect insertion followed by relaxation or by extracting environments around atoms identified by a trusted low-temperature reference procedure; bulk-crystal environments are sampled far from defect cores to ensure clean class separation. Because point defects are intrinsically sparse, defect environments are aggregated across many independent low-temperature snapshots, while bulk environments are sampled from a representative reference configuration. Each atomic environment is treated as an independent sample described by its local descriptors, and the resulting dataset is partitioned into training and validation subsets at the level of atomic environments, appropriate because the descriptor encodes local geometry independently per atom. This construction reflects the practical setting in which reliable labels are available only at low temperature, while the learned model must generalize to thermally perturbed environments.

\subsection{Denoising backbone}
\label{subsec:denoiser}

We model a thermally perturbed environment as
\begin{equation}
    x = x_0 + \epsilon,
\end{equation}
where $x_0$ lies near a low-dimensional manifold of structurally meaningful environments and $\epsilon$ is a stochastic displacement induced by thermal motion. The denoiser $D_\theta: \mathcal{X} \rightarrow \mathcal{X}$ predicts a cleaner configuration $\tilde{x}=D_\theta(x)$ from a noisy input.

We adopt the score-based denoising framework of Hsu \emph{et al.}~\cite{hsu2024scorebased} as our input-stage component: $D_\theta$ is trained on synthetically perturbed reference structures using a reconstruction objective of the form
\begin{equation}
    \mathcal{L}_{\mathrm{denoise}}(\theta)
    =
    \mathbb{E}_{x_0 \sim p_s,\,\epsilon \sim \mathcal{N}(0,\sigma^2 I)}
    \left[\lVert D_\theta(x_0+\epsilon)-x_0 \rVert_2^2\right],
\end{equation}
implemented as an equivariant graph neural network~\cite{batzner2022nequip,batatia2022mace} (so that predicted coordinate corrections transform properly under translation and rotation) and applied iteratively for a small fixed number of steps at inference. We follow the architecture and training procedure of Ref.~\cite{hsu2024scorebased} and train the denoiser on low-temperature reference structures of the systems under study, including defect-containing references; Appendix~\ref{appendix:SIA_collapse} illustrates why the training references must include defects, using the interstitial dumbbell as a case study. The denoiser is not a contribution of the present paper: we treat it as the input-level alignment module and build the cross-temperature label-transfer machinery on top of it.

\begin{remark}
Denoised structures are used here as analysis aids rather than as claims about the exact thermodynamic microstate: denoising is best interpreted as a geometric projection toward a cleaner structural representation, not as a substitute for free-energy minimization~\cite{hsu2024scorebased}. Empirically, the denoiser is iterative and stochastic, and its output exhibits run-to-run variability; this is one source of the residual cross-temperature gap that the downstream alignment terms are designed to absorb.
\end{remark}

\subsection{ACE representation and classifier}
\label{subsec:classifier}

After denoising, each local environment is embedded using the Atomic Cluster Expansion (ACE),
\begin{equation}
    \phi_{\mathrm{ACE}}: \mathcal{X} \rightarrow \mathbb{R}^{d},
\end{equation}
which provides a systematic invariant representation of local atomic geometry~\cite{drautz2019ace,drautz2019erratum}, closely related to SOAP~\cite{bartok2013representing} and atom-centered symmetry functions~\cite{behler2007generalized}. ACE is chosen over hand-tuned geometric features because it provides a complete, rotation- and permutation-invariant basis that retains sensitivity to defect topology even when residual thermal distortion remains after denoising. For environment $x$, the descriptor vector is
\begin{equation}
    z = \phi_{\mathrm{ACE}}(D_\theta(x)).
\end{equation}

The ACE vector is passed through an encoder $f_\psi$ and classifier $g_\omega$,
\begin{equation}
    f_\psi: \mathbb{R}^{d} \rightarrow \mathbb{R}^{m},
    \qquad
    g_\omega: \mathbb{R}^{m} \rightarrow \Delta^{|\mathcal{Y}|},
\end{equation}
where $\Delta^{|\mathcal{Y}|}$ denotes the probability simplex over classes. The predicted class probabilities are
\begin{equation}
    \hat{p}(y\mid x)=g_\omega\bigl(f_\psi(\phi_{\mathrm{ACE}}(D_\theta(x)))\bigr).
\end{equation}

The main supervised objective is the source-domain cross-entropy,
\begin{equation}
    \mathcal{L}_{\mathrm{task}}
    =
    -\mathbb{E}_{(x,y)\sim \mathcal{D}_s}\log \hat{p}(y\mid x).
    \label{eq:task}
\end{equation}
During training, synthetic coordinate perturbations are injected into the source environments so that the classifier sees a broader distribution than the exact low-temperature manifold alone.

\subsection{Representation-level alignment via cross-temperature contrastive learning}
\label{subsec:contrastive}

Because the target domain differs from the source domain primarily through thermal distortion, it is natural to encourage low- and high-temperature views of the same underlying atomistic region to remain close in latent space~\cite{kang2019can,park2020joint,shen2022connect}.

\paragraph{Pairing assumption.}
We use \emph{atom-wise} positive pairs: for the same simulated system, the $i$-th atom in a low-temperature snapshot and the $i$-th atom in a high-temperature snapshot form a positive pair, on the grounds that thermal evolution from the same initial cell preserves identity over the timescales we analyze. This is the natural setup when the high-temperature configuration is an MD continuation of the low-temperature one, which holds for our experiments. It is not required for the framework as a whole: the denoiser and morphology regularizer do not depend on any pairing, and feature-distribution alignment~\cite{long2015dan,sun2016deepcoral} or domain-adversarial training~\cite{ganin2015dann} can be substituted when atom-level correspondences are unreliable (e.g., after large-scale diffusion or vacancy migration).

\paragraph{Loss.}
For matched source--target pairs $\mathcal{P}=\{(i,j)\}$, we introduce a projection head $h_\eta$ and normalized embeddings,
\begin{equation}
    \bar{q}_i^s = \frac{h_\eta(f_\psi(z_i^s))}{\|h_\eta(f_\psi(z_i^s))\|_2},
    \qquad
    \bar{q}_j^t = \frac{h_\eta(f_\psi(z_j^t))}{\|h_\eta(f_\psi(z_j^t))\|_2}.
\end{equation}
For a minibatch of $B$ matched pairs, similarities are defined by
\begin{equation}
    s_{ab}=\frac{\langle \bar{q}_{i_a}^s,\bar{q}_{j_b}^t\rangle}{\tau},
\end{equation}
where $B$ is the number of matched pairs in the minibatch and $\tau$ is a fixed temperature (sharpness) hyperparameter. We train with minibatches of 200 atomic environments, so $B$ varies from batch to batch with the number of paired-snapshot atoms present ($B \le 200$), and we use $\tau = 0.05$.
We use a symmetric InfoNCE-style objective~\cite{oord2018cpc,chen2020simclr,khosla2020supcon}: the contribution of pair $a$ is
\begin{equation}
    \ell_a
    =
    -\frac{1}{2}\left[
    \log\frac{\exp(s_{aa})}{\sum_b \exp(s_{ab})}
    +
    \log\frac{\exp(s_{aa})}{\sum_b \exp(s_{ba})}
    \right],
    \label{eq:pair_ctr}
\end{equation}
where the two terms score the same positive pair against the negatives of the source-to-target and target-to-source directions, respectively, and the unweighted contrastive loss is the minibatch average
\begin{equation}
    \mathcal{L}_{\mathrm{ctr}}=\frac{1}{B}\sum_a \ell_a.
\end{equation}

\paragraph{Reliability weighting.}
Not every pair carries the same amount of useful signal. Near defect cores, or after substantial atomic motion, the atom-wise pairing assumption is weakest, and forcing those pairs into close embedding agreement can be counterproductive. We therefore softly down-weight pairs whose normalized embeddings are already far apart in projection space, on the heuristic that such pairs are likely either mislabeled (broken correspondence) or genuinely hard, and in either case should not dominate the objective. Each pair receives a reliability weight
\begin{equation}
    w_a = \exp\left(-\|\bar{q}_{i_a}^s-\bar{q}_{j_a}^t\|_2\right),
\end{equation}
and we form a normalized weighted contrastive loss,
\begin{equation}
    \mathcal{L}_{\mathrm{ctr}}^{\mathrm{w}}
    =
    \frac{1}{B}\sum_a \tilde{w}_a \ell_a,
    \label{eq:weighted_ctr}
\end{equation}
where $\ell_a$ is the per-pair contribution of Eq.~\eqref{eq:pair_ctr} and the $\tilde{w}_a = w_a / \bigl(\tfrac{1}{B}\sum_b w_b\bigr)$ are the reliability weights rescaled to unit mean over the minibatch; the weights are computed with stop-gradient, so they modulate the loss but are not themselves optimized.
The intent is not that every atom-wise correspondence be exploited, but that \emph{enough} reliable pairs are exploited to learn a temperature-robust encoder; unreliable pairs contribute little and do not corrupt training. This term therefore acts as a representation-level alignment mechanism in the spirit of feature-space domain alignment~\cite{long2015dan,sun2016deepcoral,kang2019can,park2020joint,shen2022connect}: it does not supervise target labels directly, but reduces the discrepancy between source and target latent geometry.

\subsection{Morphology-aware regularization}
\label{subsec:physics}

A classifier that is locally accurate but spatially erratic is of limited use for defect analysis: genuine point defects and defect-affected neighborhoods should appear as coherent, weakly fragmented regions rather than diffuse fields of isolated positives. We therefore add a morphology-aware regularization term that acts directly on the predicted defect probability field in the target domain. This term serves two roles simultaneously: it improves the geometric quality of the predicted field at the level of output shape, and it acts as a domain-alignment mechanism by steering target-domain predictions toward the structural regime occupied by physically valid defect fields. Because the model is trained end-to-end, the constraint also propagates upstream through the shared encoder and can improve the learned representation itself (Fig.~\ref{fig:latent}).

Importantly, the regularizer is a \emph{tiebreaker} among classifiers consistent with the supervised source loss, not a substitute for it: among predictions with comparably low task loss, it prefers the spatially coherent one.

Let $d_i \in [0,1]$ denote the predicted defect probability at atom $i$, and let $\mathcal{G}$ be a $k$-nearest-neighbor graph on atomic positions. We define a morphology score $S_{\mathrm{morph}}$ by aggregating differentiable spatial criteria computed on this graph. In the most general implementation, these criteria include graph coherence, cluster persistence, boundary consistency, and geometric compactness, with cluster-shape priors expressed as eigenvalue functionals of the defect covariance (round here; line- or plane-shaped variants, appropriate for dislocation-like and planar defects, follow by changing the eigenvalue ratio). In the present experiments we focus the training regularization on the component most directly targeting the dominant failure mode of high-temperature predictions: cluster persistence ($S_{\mathrm{pers}}$), defined in Appendix~\ref{appendix:plausibility}, whose two sub-terms enter with equal weight. The first sub-term, fragmentation, encourages predicted defect mass to be supported by persistent local connectivity rather than dispersing across isolated singletons. The second, concentration, encourages that mass to remain concentrated in a small number of coherent regions; we emphasize ``few'' rather than ``one,'' so the prior accommodates configurations with several well-separated defects rather than collapsing every prediction to a single blob. Together these two ingredients penalize the most common failure mode in high-temperature predictions, namely spatially fragmented defect fields.

We write the aggregate morphology score as
\begin{equation}
    S_{\mathrm{morph}} = \frac{\sum_m w_m S_m}{\sum_m w_m},
    \label{eq:smorph}
\end{equation}
where the $S_m$ are normalized morphology components (discussed further below) and the $w_m$ are nonnegative weights. The corresponding penalty is
\begin{equation}
    \mathcal{L}_{\mathrm{morph}} = 1 - S_{\mathrm{morph}}.
\end{equation}
We also regularize the global predicted defect count. With $M=\sum_i d_i$ the predicted defect mass, $N$ the number of atoms, and $m_0 = N\rho^*$ the expected count at a prior defect fraction $\rho^*$,
\begin{equation}
    \mathcal{L}_{\mathrm{mass}} = \frac{(M - m_0)^2}{m_0(1-\rho^*)},
\end{equation}
a squared deviation normalized by the binomial variance of the count at the prior fraction; $\rho^*$ is a target defect fraction or a weak prior estimated from the training data for the corresponding condition.

Because these penalties are evaluated on target-domain predictions rather than labeled source examples, they do more than regularize: they guide the high-temperature output distribution toward the regime occupied by physically valid defect fields, reducing source--target mismatch at the prediction level, and, through end-to-end optimization, improve feature-space alignment as well.

The complete training objective is
\begin{equation}
    \mathcal{L}
    =
    \mathcal{L}_{\mathrm{task}}
    + \lambda_{\mathrm{ctr}}\mathcal{L}_{\mathrm{ctr}}^{\mathrm{w}}
    + \lambda_{\mathrm{reg}}\bigl(\mathcal{L}_{\mathrm{morph}} + \lambda_{\mathrm{mass}}\mathcal{L}_{\mathrm{mass}}\bigr).
    \label{eq:full_loss}
\end{equation}
Equation~\eqref{eq:full_loss} is a multi-level domain-alignment objective, as discussed previously.
The set of morphology terms $\{S_m\}$ and their weights are user-customizable (our implementation supports coherence, boundary-consistency, and compactness terms beyond the persistence term used here), so the framework can be specialized to other defect morphologies without changing the rest of the pipeline. Full definitions are given in Appendix~\ref{appendix:plausibility}.

\subsection{Training algorithm and inference}
\label{subsec:algorithm}

ACE descriptors for source and target snapshots are precomputed after denoising and reused during classifier training for efficiency. Training proceeds on labeled source batches together with unlabeled target-domain batches used only for alignment. To stabilize optimization, we first train with the supervised source loss alone and activate the contrastive term after a short warm-up. The weights of the non-supervised losses ($\lambda_{\mathrm{ctr}}$, $\lambda_{\mathrm{reg}}$, $\lambda_{\mathrm{mass}}$) are adjusted automatically during training using a gradient-norm-based heuristic with exponential moving averages: at each step, the gradient norm of each auxiliary loss is compared with that of the supervised loss, and the auxiliary weight is rescaled so that its contribution to the total gradient stays a fixed (small) fraction of the supervised contribution. This keeps the auxiliary terms balanced relative to the classification objective and reduces manual retuning across systems.

\begin{algorithm}[t]
\caption{Training with multi-level domain alignment}
\begin{algorithmic}[1]
\State Precompute denoised structures and ACE descriptors for labeled source snapshots and unlabeled target-domain snapshots
\For{each training epoch}
    \For{each minibatch}
        \State Sample labeled source atoms and unlabeled target-domain atoms
        \State Compute latent embeddings with $f_\psi$
        \State Evaluate supervised loss $\mathcal{L}_{\mathrm{task}}$ on the source labels
        \If{correspondence information is available}
            \State Build matched source--target pairs $\mathcal{P}$
        \EndIf
        \If{warm-up is complete}
            \State Compute weighted contrastive loss $\mathcal{L}_{\mathrm{ctr}}^{\mathrm{w}}$
        \Else
            \State Set $\mathcal{L}_{\mathrm{ctr}}^{\mathrm{w}} = 0$
        \EndIf
        \State Compute per-atom defect probabilities $\{d_i\}$ on the target-domain batch
        \State Evaluate $\mathcal{L}_{\mathrm{morph}}$ and $\mathcal{L}_{\mathrm{mass}}$
        \State Update the weights of $\mathcal{L}_{\mathrm{ctr}}^{\mathrm{w}}$, $\mathcal{L}_{\mathrm{morph}}$, and $\mathcal{L}_{\mathrm{mass}}$ via gradient-norm balancing with EMA
        \State Update parameters using Eq.~\eqref{eq:full_loss}
    \EndFor
\EndFor
\end{algorithmic}
\label{alg:training}
\end{algorithm}

At inference, each high-temperature snapshot is processed as follows:
\begin{enumerate}[leftmargin=1.4em]
    \item apply the denoiser to atomic coordinates;
    \item compute ACE descriptors on denoised local environments;
    \item obtain latent representations and class probabilities;
    \item extract per-atom defect probabilities or hard class assignments; and
    \item optionally apply a light spatial refinement step to remove isolated spurious detections.
\end{enumerate}
This produces both interpretable probability fields and discrete defect predictions suitable for visualization or downstream analysis.

\subsection{A compact generalization perspective}
\label{subsec:theory}

The purpose of the denoiser and the alignment regularizers is intuitive: they reduce the discrepancy between the low-temperature source domain and the high-temperature target domain, in the spirit of classical domain-adaptation analyses~\cite{bendavid2010theory}. A concise way to express this is through the target risk
\begin{equation}
    R_t(h)=\mathbb{E}_{(x,y)\sim p_t}\,\ell(h(x),y),
\end{equation}
where $\ell(\cdot,\cdot)$ is a bounded per-sample classification loss (e.g., the cross-entropy of Eq.~\eqref{eq:task}), for the composite predictor
\begin{equation}
    h(x)=g_\omega\bigl(f_\psi(\phi_{\mathrm{ACE}}(D_\theta(x)))\bigr).
\end{equation}
Under standard domain-adaptation assumptions~\cite{bendavid2010theory,ganin2015dann}, one expects a bound of the schematic form
\begin{equation}
    R_t(h) \lesssim R_s(h) + \mathrm{disc}(z_\#p_s,z_\#p_t) + \text{model mismatch},
\end{equation}
where $z_\#p$ denotes the distribution of latent representations obtained by applying $z$ to inputs sampled from $p$ and $\mathrm{disc}(\cdot,\cdot)$ is a suitable measure of discrepancy between the source and target representation distributions~\cite{long2015dan,sun2016deepcoral}. The three components act at different but coupled levels: denoising reduces discrepancy already in input space, contrastive learning reduces it in representation space~\cite{kang2019can,park2020joint,shen2022connect}, and the morphology regularizer constrains target-domain predictions while indirectly improving representation alignment through the shared encoder.

\subsection{Simulation protocol and implementation details}
\label{subsec:setup}

Molecular dynamics simulations were performed using the LAMMPS package~\cite{plimpton1995fast,thompson2022lammps}. For the point-defect studies, atomic interactions in Fe were modeled using an embedded-atom method (EAM)~\cite{daw1984eam} potential developed by Zhou \emph{et al.}~\cite{zhou2018fe}, which enables stable molecular dynamics simulations at high temperatures across all three crystal structures (FCC, BCC, and HCP). The melting temperature for all crystal structures is $T_m\approx2300$~K. Fe was chosen for these studies because it admits all three parent crystal classes within a single chemistry, isolating the effect of cross-temperature transfer from chemistry-specific confounds; the framework itself does not depend on any Fe-specific feature, and the grain-boundary study described below applies it to Al using the EAM potential of Mishin \emph{et al.}~\cite{mishin1999al} with a melting temperature of $T_m\approx1040$~K.

For the point-defect simulations, simulations were carried out in the isothermal--isobaric (NPT) ensemble using a Nose--Hoover thermostat and barostat, with a 1\,fs timestep and periodic boundary conditions. Each simulation cell contained approximately 4000 atoms. For each configuration, the system was equilibrated for 3000 timesteps, after which it reached a stable thermodynamic state, and the equilibrated snapshot was used for analysis. Snapshots were generated across a temperature range spanning low-temperature conditions to near-melting regimes for all three crystal structures. Point defects were introduced by removing atoms (vacancy-type configurations) or inserting additional atoms followed by short equilibration (SIA configurations). The training dataset was constructed from low-temperature (energy minimized) snapshots, where structural labels are reliable: because point defects are sparse, defect environments were aggregated across 140 independently seeded snapshots, yielding on the order of $10^3$ defect samples, while bulk environments were sampled from a representative reference snapshot. The dataset was randomly partitioned into training and validation subsets at the level of atomic environments, and evaluation used separate high-temperature snapshots that were never seen during training, ensuring that all reported results reflect generalization across temperature rather than memorization. Reported uncertainties are standard deviations across independent runs.

The large-scale validation of Section~\ref{subsec:bigsim} uses a one-million-atom FCC Fe simulation at $T/T_m=0.91$ under 3\,GPa hydrostatic tension, initialized with a dilute vacancy population and one thousand inserted SIAs, with analyzed frames spanning 10\,ps to 2.49\,ns. Ground truth was constructed by Wigner--Seitz occupancy analysis~\cite{nordlund1998defect} on the denoised coordinates with atom identities tracked across frames, defining vacancy and interstitial sites independently of the learned pipeline.

The GB study of Section~\ref{subsec:gb} uses Al bicrystals containing two distinct boundaries. The $\Sigma5(3\,1\,0)$ GB with crystals oriented as $[0\,0\,1] \parallel [0\,0\,1]$, $[3\,1\,0] \parallel [3\,\bar{1}\,0]$, and $[\bar{1}\,3\,0] \parallel [1\,3\,0]$ was simulated at 900\,K with an applied tensile stress of about 4.8~GPa by adding forces to surface atoms. The $\Sigma11(2\,1\,1)$ GB with crystals oriented as $[4\,\bar{1}\,\bar{7}] \parallel [4\,\bar{7}\,\bar{1}]$, $[2\,1\,1] \parallel [2\,1\,1]$, and $[2\,\bar{6}\,2] \parallel [\bar{2}\,\bar{2}\,6]$ was simulated at 300\,K and no applied stress. The two competing GB phases/complexions are defined by fully minimized reference structures, and boundary atoms (identified by PTM) are classified between the phases by an MLP trained on those references; validation uses grouped leave-one-out cross-validation at the level of snapshots or noise realizations.
GB phase/complexion transitions were induced by varying the temperature and/or applied normal stress, depending on the GB behavior.

All learned models were implemented in PyTorch~\cite{paszke2019pytorch}. The denoiser followed the reference implementation of Ref.~\cite{hsu2024scorebased} with default hyperparameters and was applied for 8 iterations at inference. Denoising was performed on a GPU and required approximately 3 hours for the full cross-temperature benchmark dataset (all systems, temperatures, and runs); the ACE-based classifier trained on a CPU within approximately 2 minutes, highlighting the lightweight nature of the classifier once descriptor features are computed.

\subsection{Baselines}
\label{subsec:baselines}

We compare the proposed pipeline against a ladder of configurations of increasing sophistication, each isolating one level of the framework:

\begin{enumerate}[leftmargin=1.4em]
    \item \textbf{PTM:}~\cite{larsen2016ptm} a geometry-based reference method for local structure assignment, included as a widely used classical baseline.
    \item \textbf{ACE + MLP:} a supervised descriptor-based classifier~\cite{drautz2019ace} trained on low-temperature data and applied directly to raw high-temperature configurations, without denoising or alignment. This represents naive cross-temperature transfer, exposed to the full domain shift.
    \item \textbf{+ denoiser:}~\cite{hsu2024scorebased} the same descriptor-based classification applied after denoising, isolating input-level alignment.
    \item \textbf{+ contrastive:} adding cross-temperature contrastive alignment in latent space~\cite{kang2019can,park2020joint}, isolating representation-level alignment.
    \item \textbf{+ morphology:} the full model, additionally including the morphology term on target-domain predictions, the deployed configuration throughout this paper.
\end{enumerate}

The ground-truth evaluation scores each configuration directly against Wigner--Seitz labels (Tables~\ref{tab:metrics} and~\ref{tab:ablation}); the label-free closeness suite additionally reports a random-permutation baseline that anchors its $[0,1]$ scale from below (Table~\ref{tab:defect_field_closeness}).

\subsection{Ground-truth evaluation via Wigner--Seitz analysis}
\label{subsec:ws_eval}

For the point-defect systems, ground truth at any temperature is constructed by Wigner--Seitz occupancy analysis~\cite{nordlund1998defect} on the denoised coordinates: a perfect-crystal site template is registered to the frame (including a fitted global offset), each atom is assigned to its nearest site, and vacancies and interstitials are read off as empty and doubly occupied sites, respectively. Per-atom defect labels follow a core-plus-halo construction with a per-lattice radius calibrated on the cold training snapshots, where the resulting labels reproduce the clean geometric labels exactly. These labels are used for evaluation only: training never sees them, nor any other high-temperature label, so the cross-temperature setting is preserved. Detection is scored at the defect level. True defect objects are defined as Wigner--Seitz site clusters of at least five atoms (smaller fragments are halo fringe), and predicted defect objects (connected clusters of flagged atoms) are matched to them one-to-one, greedily, by periodic-boundary-aware centroid distance within the clustering cutoff. Position-based matching is deliberately chosen over overlap-based rules: the Wigner--Seitz core sites pin each defect's position exactly, whereas the spatial extent of a truth object depends on the halo radius, so centroid matching is insensitive to that radius (its verdicts are stable under halo-radius perturbation, where overlap-based matching flips). Percolating clusters have no defined center and are ineligible, so degenerate flag-everything predictions can never match; in practice, matched pairs sit within about 1.5\,\AA{} of the true center, while the nearest rejected pair lies at 4.2\,\AA. We report object-level recall, precision, F1, and the count of unmatched (false) predicted objects per snapshot. The same construction, with atom identities tracked across frames, provides the ground truth for the large-scale study of Section~\ref{subsec:bigsim}.

\subsection{Label-free evaluation metrics}
\label{subsec:metrics}

Wigner--Seitz ground truth is available only when defects live on a known site registry; for defect classes without one (GB phases, and extended defects generally), no atom-wise truth exists at temperature, and evaluation needs a different foundation. Rather than scoring predictions against a single (unavailable) truth, we adopt a triangulation strategy borrowed from segmentation evaluation~\cite{zhang2008survey}: predictions are scored along five complementary axes that a good defect structure should jointly satisfy: spatial smoothness, cluster structure, geometric compactness, boundary quality, and consistency with a strong physics-based reference signal (denoiser-assisted PTM~\cite{larsen2016ptm,hsu2024scorebased}). Each axis is reported as a closeness, $1-\lvert m_{\mathrm{model}}-m_{\mathrm{ref}}\rvert$, to the reference behavior (Table~\ref{tab:defect_field_closeness}).

We treat denoiser+PTM not as a gold standard but as one triangulating signal among five: it is an independent, physics-based pipeline whose predictions are reasonable near melting, and the fifth axis reports how close a prediction is to this external reference. A model that systematically resembles denoiser+PTM will score well on that axis by construction; this is why the suite couples it with four label-free spatial axes that denoiser+PTM does not define, and why we anchor the entire evaluation against exhaustive Wigner--Seitz ground truth where it can be constructed (Tables~\ref{tab:metrics} and~\ref{tab:ablation}, Sections~\ref{subsec:bigsim} and~\ref{subsec:closeness}).

All metrics are normalized to lie in $[0,1]$, with higher values indicating better agreement with reference behavior. We report five scalar metrics: \textbf{Coherence}, \textbf{Fragmentation}, \textbf{Compact geometry}, \textbf{Boundary}, and \textbf{Physics}.

\paragraph{Coherence.}
Coherence measures the spatial smoothness of the predicted defect probability field using a graph-based criterion analogous to normalized total variation~\cite{rudin1992nonlinear,gilboa2008nonlocal}, combined with a local averaging consistency term in the spirit of spatial-autocorrelation statistics~\cite{moran1950notes}. High coherence indicates that neighboring atoms receive consistent predictions.

\paragraph{Fragmentation.}
Fragmentation evaluates whether predicted defects are concentrated into a small number of connected components rather than dispersed across many isolated regions, incorporating a concentration term that favors dominant clusters over evenly distributed small clusters. Tracking components across linking scales follows the logic of zero-dimensional persistence in topological data analysis~\cite{carlsson2009topology,edelsbrunner2010computational}, and the concentration term is an inverse participation ratio~\cite{thouless1974electrons}, equivalent to the Simpson concentration index~\cite{simpson1949measurement}.

\paragraph{Compact geometry.}
Compact geometry measures whether predicted defect clusters are spatially localized and approximately isotropic, computed from cluster shape statistics such as the radius of gyration and covariance eigenvalues, standard shape descriptors from gyration-tensor analysis of molecular conformations~\cite{theodorou1985shape}.

\paragraph{Boundary.}
Boundary consistency evaluates whether predicted defect regions form contiguous domains with well-defined interfaces, based on the fraction of neighboring atoms that disagree with the predicted defect label, analogous to boundary-quality measures in image segmentation~\cite{martin2004learning}.

\paragraph{Physics.}
Physics consistency measures agreement with the physically motivated reference signal derived from denoiser-assisted PTM analysis~\cite{larsen2016ptm,hsu2024scorebased}, combining an effect-size measure~\cite{cohen1988statistical} with a ranking metric (the area under the receiver operating characteristic curve, AUC~\cite{hanley1982meaning}) to quantify separation between defect and non-defect regions.

\paragraph{Reference-relative scoring.}
To facilitate comparison across heterogeneous criteria, all values are reported relative to a reference atomic structure and normalized to $[0,1]$. We additionally report a random baseline, obtained by scoring random permutations of each model's per-atom defect probabilities in the manner of permutation null models~\cite{good2000permutation}, which corresponds to spatially unstructured predictions and highlights that meaningful defect structures exhibit non-random spatial organization. These metrics correspond directly to the formal definitions of GCS, MCPS, MCGS, BCS, and PCS in Appendix~\ref{appendix:metrics}.

\backmatter

\bmhead{Data availability}

The data supporting the findings of this study are available from the corresponding author upon reasonable request. Upon publication, the authors intend to release the curated benchmark splits, trained model checkpoints, and scripts necessary to reproduce the main figures, subject to repository and institutional constraints.

\bmhead{Code availability}

Code for denoising, descriptor extraction, classifier training, and evaluation will be made available upon publication or can be shared by the corresponding author upon reasonable request during review.

\bmhead{Acknowledgements}

This material is based upon work supported by the National Science Foundation under Grant No. 2409835. 
YF gratefully acknowledges support from the Rutgers AI and Data Science Collaboratory.
The authors thank Dr. Alexander Stukowski and Dr. Daniel Utt for helpful discussions.

\section*{Declarations}

\textbf{Funding} National Science Foundation under Grant No. 2409835

\textbf{Conflict of interest/Competing interests} RBS is Editor-in-Chief of the Journal of Materials Science: Materials Theory, a Springer Nature journal.

\textbf{Ethics approval and consent to participate} Not applicable.

\textbf{Consent for publication} Not applicable.

\textbf{Materials availability} Not applicable.

\textbf{Author contributions} YF: conceptualization, data curation, formal analysis, investigation, methodology, software, validation, visualization, writing - original draft; JK: formal analysis, methodology, writing - review \& editing; QQZ: conceptualization, data curation, formal analysis, methodology, software, writing - review \& editing; PB: formal analysis, investigation, methodology, writing - review \& editing; JMG: data curation, investigation, writing - review \& editing; RBS: conceptualization, funding acquisition, investigation, methodology, project administration, supervision, writing - review \& editing; AAE: conceptualization, funding acquisition, investigation, methodology, project administration, supervision, writing - review \& editing

\begin{appendices}

\section{Spatial morphology metrics}
\label{appendix:plausibility}

This appendix defines the morphology components $S_m$ entering the aggregate score of Eq.~\eqref{eq:smorph}, each a differentiable functional of the predicted defect-probability field evaluated on a spatial graph. The implementation is intentionally modular: components can be used individually or in combination, depending on the task. In the present experiments, the training regularizer uses the cluster-persistence component $S_{\mathrm{pers}}$, with its two sub-terms equally weighted, while the broader family is retained for diagnostics and future extensions. These training-time components are the differentiable counterparts of the discrete evaluation metrics of Appendix~\ref{appendix:metrics} ($S_{\mathrm{coh}} \leftrightarrow \mathrm{GCS}$, $S_{\mathrm{pers}} \leftrightarrow \mathrm{MCPS}$).

\subsection{Spatial graph construction}

Let $\mathbf{x}_i \in \mathbb{R}^3$ denote atomic coordinates. For each neighborhood size $k$, we construct a mutual $k$-nearest-neighbor graph~\cite{luxburg2007tutorial}. For an edge $(i,j)$ with Euclidean distance $r_{ij}=\|\mathbf{x}_i-\mathbf{x}_j\|$, we define a Gaussian edge weight
\begin{equation}
    w_{ij}^{(k)} = \exp\left[-\left(\frac{r_{ij}}{\sigma_k}\right)^2\right],
\end{equation}
where $\sigma_k$ is set to the median, over atoms, of the distance to the $k$-th nearest neighbor.

To incorporate multiple spatial scales, we also define softened thresholded weights
\begin{equation}
    \tilde{w}_{ij}^{(k,t)} =
    \operatorname{sigmoid}\!\left(\alpha\left(t-\frac{r_{ij}}{\sigma_k}\right)\right)
    w_{ij}^{(k)},
\end{equation}
where $t$ is a scale parameter and $\alpha$ controls the transition sharpness.

\subsection{Coherence component $S_{\mathrm{coh}}$}

The coherence component measures smoothness of the defect probability field, in the spirit of (nonlocal) total-variation regularization~\cite{rudin1992nonlinear,gilboa2008nonlocal}. Let $d_i \in [0,1]$ denote the predicted defect probability at node $i$. We define the normalized total variation
\begin{equation}
    \mathrm{TV}_k =
    \frac{\sum_{(i,j)} w_{ij}^{(k)} |d_i-d_j|}
         {\sum_{(i,j)} w_{ij}^{(k)}},
\end{equation}
and the neighbor-averaged field
\begin{equation}
    \bar{d}_i =
    \frac{\sum_j w_{ij}^{(k)} d_j}
         {\sum_j w_{ij}^{(k)}},
    \label{eq:nbr_avg}
\end{equation}
i.e., $\bar{d}_i$ is the weighted average of the defect probabilities over the graph neighbors of atom $i$. The coherence component at neighborhood size $k$ is
\begin{equation}
    S_{\mathrm{coh}}^{(k)}
    =
    \frac{1}{2}
    \left(
    1-\mathrm{TV}_k
    +
    1-\frac{\sum_i |d_i-\bar d_i|}{\sum_i (d_i+\bar d_i)}
    \right),
\end{equation}
with $\bar d_i$ given by Eq.~\eqref{eq:nbr_avg}. High values of $S_{\mathrm{coh}}^{(k)}$ indicate that neighboring atoms receive mutually compatible defect probabilities.

\subsection{Cluster-persistence component $S_{\mathrm{pers}}$}

To quantify whether predicted defects form coherent connected regions that persist across spatial scales (a soft analogue of tracking connected components through a scale filtration~\cite{carlsson2009topology}), we define the soft edge affinity
\begin{equation}
    a_{ij}^{(k,t)} = \tilde{w}_{ij}^{(k,t)} d_i d_j,
\end{equation}
and the corresponding node support
\begin{equation}
    A_i = \sum_j a_{ij}^{(k,t)}.
\end{equation}
A fragmentation-sensitive term is then
\begin{equation}
    S_{\mathrm{pers,frag}}^{(k,t)}
    =
    \frac{\sum_i (1-e^{-A_i}) d_i}{\sum_i d_i},
\end{equation}
which increases when defect probability mass is supported by persistent local connectivity.

To complement this, we define a concentration term. Let $c_i$ be the result of a small fixed number of damped, degree-normalized diffusion steps of the defect probabilities $d$ over the soft thresholded graph (in the spirit of label propagation~\cite{zhu2003semi}), so that $c$ represents defect mass redistributed along persistent connections, and let $M=\sum_i d_i$ denote the total defect mass. The concentration term is the mass-scaled inverse participation ratio~\cite{thouless1974electrons}
\begin{equation}
    S_{\mathrm{pers,conc}}
    =
    \min\!\left\{1,\; M\,\frac{\sum_i c_i^2}{\left(\sum_i c_i\right)^2}\right\}.
\end{equation}
Intuitively, this term is large when the predicted defect mass remains concentrated on a small number of coherent regions and falls as mass diffuses across many weakly connected islands.

The resulting cluster-persistence component is
\begin{equation}
    S_{\mathrm{pers}}
    =
    \frac{1}{2}\left(
    S_{\mathrm{pers,frag}}
    +
    S_{\mathrm{pers,conc}}
    \right).
\end{equation}

\subsection{Additional implemented components}

For completeness, the implementation also supports additional morphology components $S_m$ beyond the two emphasized during training in the present experiments.

\paragraph{Boundary-consistency component.}
A boundary-consistency component measures whether predicted defect regions are surrounded by compatible local neighborhoods rather than dominated by isolated boundary points.

\paragraph{Geometric-compactness component.}
A geometric-compactness component evaluates whether the predicted defect field remains spatially localized, using differentiable surrogates for compactness and shape regularity.

\paragraph{Shape priors via eigenvalue functionals.}
Cluster-shape priors enter only through an eigenvalue functional of the (differentiable) defect covariance. The present experiments use the isotropic ratio $\lambda_{\min}/\lambda_{\max}$, which favors round clusters; line- and plane-shaped priors (appropriate for dislocation-like and planar defects) correspond to different eigenvalue ratios of the same covariance, with all other machinery unchanged.

\subsection{Aggregation across scales}

For each morphology component $S_m$, values are computed across multiple neighborhood sizes $k$ and, where applicable, multiple scale parameters $t$. The component entering Eq.~\eqref{eq:smorph} is obtained by averaging or weighted averaging over the selected values of $k$ and $t$ (e.g., $S_{\mathrm{coh}}$ averages $S_{\mathrm{coh}}^{(k)}$ over $k$). This multi-scale aggregation reduces sensitivity to a single neighborhood choice and improves robustness in practice.

\section{Detailed definition of evaluation metrics}
\label{appendix:metrics}

This appendix defines the five evaluation metrics (GCS, MCPS, MCGS, BCS, PCS) reported in the main text. They are discrete, evaluation-time counterparts of the differentiable training components of Appendix~\ref{appendix:plausibility}. Let $G=(V,E)$ denote the mutual $k$-nearest-neighbor graph on atomic positions, with interatomic distances $r_{ij}$ computed under the periodic minimum-image convention and Gaussian edge weights $w_{ij}=\exp[-(r_{ij}/\sigma_k)^2]$, where $\sigma_k$ is the median of all $k$-nearest-neighbor distances in the frame. Let $d_i \in [0,1]$ denote the predicted defect probability at node $i$.

\subsection{Defect set construction}

Because the predicted defect field is probabilistic, we define a discrete defect set by selecting a number of atoms consistent with the predicted total defect mass. Specifically, let
\begin{equation}
    m = \max\left\{1, \mathrm{round}\left(\sum_i d_i\right)\right\}.
\end{equation}
The defect set $D$ is then defined as the set of the top-$m$ atoms ranked by $d_i$. This construction yields a consistent defect fraction across predictions without requiring an externally fixed threshold.

\subsection{Spatial coherence}

We quantify smoothness using an edge-weighted normalized total variation~\cite{rudin1992nonlinear,gilboa2008nonlocal}:
\begin{equation}
    \mathrm{TV}(d) = \frac{\sum_{(i,j)\in E} w_{ij}\,|d_i - d_j|}{\sum_{(i,j)\in E} w_{ij}}.
\end{equation}
We also compute a local averaging consistency term,
\begin{equation}
    \mathrm{BC}(d) = 1 - \frac{\sum_i |d_i - \bar{d}_i|}{\sum_i \left(d_i + \bar{d}_i\right)},
\end{equation}
where $\bar{d}_i$ is the weighted neighbor average of Eq.~\eqref{eq:nbr_avg}. The global coherence score is
\begin{equation}
    \mathrm{GCS} = \frac{1}{2}\left(1 - \mathrm{TV}(d) + \mathrm{BC}(d)\right);
\end{equation}
it is the evaluation-time counterpart of $S_{\mathrm{coh}}$, computed with the same functional form on the same weighted graph.

\subsection{Boundary consistency}

For the defect set $D$, we define for each defect atom the edge-weighted fraction of its graph neighbors lying outside $D$:
\begin{equation}
    b_i = \frac{\sum_{j \in \mathcal{N}(i),\, j \notin D} w_{ij}}{\sum_{j \in \mathcal{N}(i)} w_{ij}},
\end{equation}
with the convention $b_i = 1$ for defect atoms with no graph neighbors, so that isolated predictions cannot raise the score. The boundary consistency score is
\begin{equation}
    \mathrm{BCS} = 1 - \frac{1}{|D|} \sum_{i \in D} b_i.
\end{equation}

\subsection{Cluster structure}

Let $\{C_1, \dots, C_K\}$ denote the connected components of the thresholded defect subgraph, whose vertices are the atoms of $D$ and whose edges are the graph edges between them with $r_{ij} \le t\,\sigma_k$, where $t$ is a cluster-linking scale swept as described below; counting components across this sweep follows the logic of zero-dimensional persistence~\cite{carlsson2009topology,edelsbrunner2010computational}. We define fragmentation as
\begin{equation}
    \text{Fragmentation} = 1 - \frac{K - 1}{|D| - 1},
\end{equation}
and concentration as the Simpson--participation index~\cite{simpson1949measurement,thouless1974electrons} of the component-size distribution,
\begin{equation}
    \text{Concentration} = \sum_{i=1}^K \left(\frac{|C_i|}{|D|}\right)^2.
\end{equation}
These are combined into a cluster structure score:
\begin{equation}
    \mathrm{MCPS} = \frac{1}{2}(\text{Fragmentation} + \text{Concentration}).
\end{equation}

\subsection{Geometric compactness}

For each cluster $C$, let
\begin{equation}
    \Sigma_C = \frac{1}{\lvert C \rvert} \sum_{i \in C}
    (\mathbf{x}_i - \bar{\mathbf{x}}_C)(\mathbf{x}_i - \bar{\mathbf{x}}_C)^{\top}
\end{equation}
denote the covariance matrix of the atomic positions in the cluster, where $\bar{\mathbf{x}}_C$ is the cluster centroid. Compactness compares the cluster's radius of gyration $r_g = \sqrt{\operatorname{tr}\Sigma_C}$ with that of an ideally close-packed cluster of the same size at the local atomic spacing, $r_g^{\mathrm{ideal}} = 0.43\,\sigma_k\,|C|^{1/3}$:
\begin{equation}
    \text{compactness}(C) = \min\left\{1,\; \frac{r_g^{\mathrm{ideal}}}{r_g}\right\}.
\end{equation}
Sphericity is the ratio of the smallest to largest eigenvalue of $\Sigma_C$ (a standard gyration-tensor shape descriptor~\cite{theodorou1985shape}), computed for clusters of at least five atoms and set to zero otherwise. Cluster scores are averaged with weights $|C|/|D|$ (each cluster's share of the defect set), and the two axes combine as
\begin{equation}
    \mathrm{MCGS} = \frac{1}{2}(\text{compactness} + \text{sphericity}).
\end{equation}

\subsection{Physics consistency}

Let $s_i$ denote a physically motivated reference signal derived from denoiser-assisted PTM analysis~\cite{larsen2016ptm,hsu2024scorebased}. We compute the absolute standardized mean difference (Cohen's $d$~\cite{cohen1988statistical}) between the defect set and its complement,
\begin{equation}
    E =
    \frac{\left|\,\mathbb E[s_i \mid i \in D] - \mathbb E[s_i \mid i \in \bar D]\,\right|}{s_p},
\end{equation}
where $s_p$ is the pooled standard deviation of the two groups, together with an orientation-independent ranking score $A = \max(\mathrm{AUC},\, 1-\mathrm{AUC})$, where AUC is the area under the receiver operating characteristic curve~\cite{hanley1982meaning} for separating $D$ from $\bar D$ by $s$, computed via its rank-sum (Mann--Whitney) equivalence~\cite{mann1947test}. The resulting physics consistency score,
\begin{equation}
    \mathrm{PCS}
    =
    \frac{1}{2}\left(
    \tanh(E)
    +
    (2A - 1)
    \right),
\end{equation}
lies in $[0,1]$ by construction.

\subsection{Multi-scale aggregation}

All graph-based metrics are computed for each neighborhood size $k \in \{8, 10, 12, 14, 16\}$. The component-based scores (fragmentation, concentration, compactness, sphericity) are additionally computed over the cluster-linking sweep $t \in \{1.1, 1.2, 1.3, 1.4\}$ (in units of $\sigma_k$) and averaged across $t$ within each $k$. Each score is then summarized by its median across the five neighborhood sizes, with sub-scores aggregated before pairwise combination (e.g., MCPS is the mean of the median-aggregated fragmentation and concentration scores). The median across neighborhood sizes reduces sensitivity to any single graph-construction choice.

\subsection{Reference-relative normalization}

For each metric $M \in \{\mathrm{GCS}, \mathrm{MCPS}, \mathrm{MCGS}, \mathrm{BCS}, \mathrm{PCS}\}$, let $m_{\mathrm{ref}}$ denote its value computed on the reference defect structure and $m_{\mathrm{model}}$ its value computed on the model prediction. We report the closeness
\begin{equation}
    \mathrm{Score}(M) = 1 - \lvert m_{\mathrm{model}} - m_{\mathrm{ref}} \rvert,
\end{equation}
matching the notation of the main text and mapping heterogeneous metrics to a common $[0,1]$ scale for side-by-side comparison.

\section{Collapse of SIA by denoiser}
\label{appendix:SIA_collapse}

\begin{figure}[htbp]
\centering
\includegraphics[width=\linewidth]{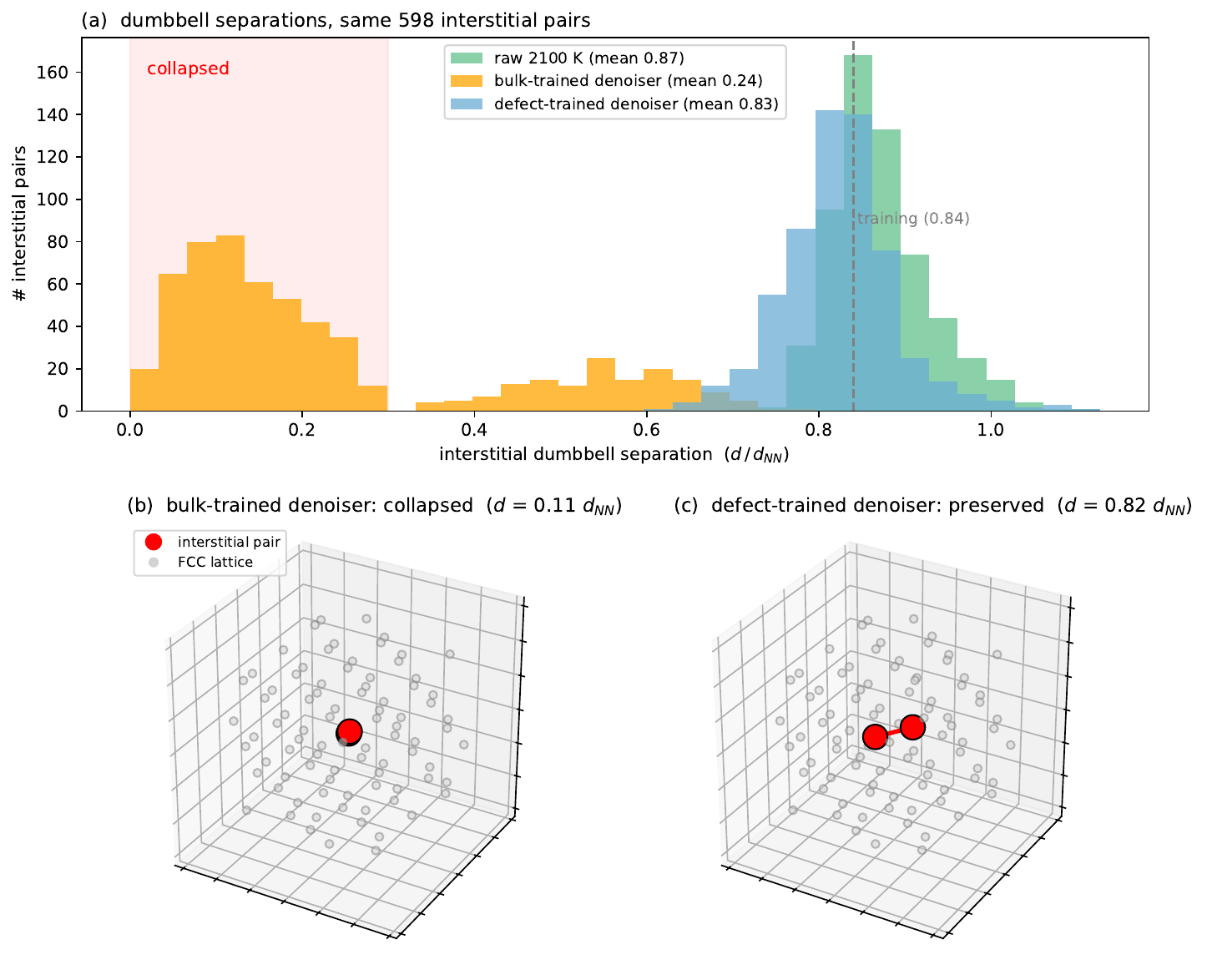}
\caption{\textbf{A denoiser trained only on perfect crystals erases
interstitials; retraining it with defect examples fixes this.}
\textbf{(a)}~Separation $d$ of the two atoms forming each interstitial
dumbbell, in units of the nearest-neighbor distance $d_\mathrm{NN}$, for the
same 598 interstitial pairs identified by Wigner--Seitz analysis of the
1M-atom FCC Fe test frame ($T/T_m = 0.91$, 3\,GPa). In the raw MD frame the pairs
sit near $d = 0.87\,d_\mathrm{NN}$ (green), matching the low-temperature
training geometry (dashed line, 0.84). The crystal-trained denoiser reads the
two-atoms-on-one-site compression as noise and merges each pair: the
distribution collapses to a mean of $0.24\,d_\mathrm{NN}$ (orange), largely
below $0.3\,d_\mathrm{NN}$ (shaded), where the pair is effectively a single
atom on a normal lattice site. The defect-trained denoiser restores the
physical geometry (blue, mean $0.83\,d_\mathrm{NN}$).
\textbf{(b,\,c)}~One representative pair (raw separation
$0.83\,d_\mathrm{NN}$) after each denoiser, with the surrounding FCC lattice
in gray. The crystal-trained model fuses the two atoms into what appears to be
one (b, $d = 0.11\,d_\mathrm{NN}$); the defect-trained model preserves the
dumbbell (c, $d = 0.82\,d_\mathrm{NN}$). The collapse in (b) is what made
SIAs invisible to the downstream classifier; all classification
results in this work use the defect-trained denoiser.}
\label{fig:SIA_collapse}
\end{figure}

When testing the pre-trained denoiser developed by Hsu \emph{et al.}~\cite{hsu2024scorebased}, we observed unphysical collapse of SIAs, as shown in Figure~\ref{fig:SIA_collapse}.
This is because the pre-trained denoiser was only trained on perfect crystals, and attempted to locate all atoms onto lattice sites.
This is manifestly unphysical in the case of SIAs.
This observation motivates training structure-specific denoisers for all atomic structures of interest.

\end{appendices}

\bibliography{references}

\end{document}